\documentclass[twocolumn]{28onr_art}
\usepackage{ulem}
\usepackage[comma]{natbib}
\usepackage{graphicx}
\usepackage[intlimits]{amsmath}

\newcommand\Rey{\mbox{\textit{Re}}}  
\newcommand\Fro{\mbox{\textit{Fr}}}  

\def\be{\begin{eqnarray}}
\def\ee{\end{eqnarray}}
\def\benl{\begin{eqnarray*}}
\def\eenl{\end{eqnarray*}}

\newcommand{\nwc}{\newcommand}
\nwc{\bm}{\boldmath}
\nwc{\m}{\mbox}
\nwc{\ubm}{\unboldmath}
\nwc{\bmU}{\m{\bm$U$\ubm}}
\nwc{\bmX}{\m{\bm$X$\ubm}}
\nwc{\bmu}{\m{\bm$u$\ubm}}
\nwc{\bmx}{\m{\bm$x$\ubm}}
\nwc{\bmz}{\m{\bm$z$\ubm}}
\nwc{\bmv}{\m{\bm$v$\ubm}}
\nwc{\bmw}{\m{\bm$w$\ubm}}
\nwc{\bmW}{\m{\bm$W$\ubm}}
\nwc{\bmn}{\m{\bm$n$\ubm}}
\nwc{\bmG}{\m{\bm$G$\ubm}}
\nwc{\bmF}{\m{\bm$F$\ubm}}
\nwc{\bmI}{\m{\bm$I$\ubm}}
\nwc{\bmN}{\m{\bm$N$\ubm}}
\nwc{\bmP}{\m{\bm$P$\ubm}}
\nwc{\bmcalP}{\m{\bm $\cal P$\ubm}}
\nwc{\bmV}{\m{\bm$V$\ubm}}
\nwc{\bmS}{\m{\bm$S$\ubm}}

\begin{document}
\pagenumbering{arabic}

%
%
%
%
%
%
%
%
\title{Parameterization of the Near-Field Internal Wave Field Generated by a Submarine}
%
%
%
\author{James W. Rottman$^1$ , Kyle A. Brucker$^1$, Douglas Dommermuth$^1$ and Dave Broutman$^2$}
\affiliation{$^1$Science Applications International Corporation, USA \\  $^2$Computational Physics, Inc., USA}
\maketitle
%
%
\section{ABSTRACT}

We attempt to gain some insight into the modeling of the generation of
internal waves produced by submarines traveling in the littoral regions
of the ocean with the use of high fidelity numerical simulations. These
numerical simulations are shown to be capable of simulating high
Reynolds number flow around bodies, including the effects of stable
stratification.  In addition, we use the results of these detailed
numerical studies to test and revise the source distribution
parameterizations of the near-field waves that have been used in
analytical studies based on linear theory. Such parameterizations have
been shown to be useful in initializing ray-tracing schemes that can be
used for computing wave propagation through realistic oceans with
variable background properties.   For simplicity, we focus on the
idealized case of a spherical body traveling horizontally at constant
speed through a uniformly stratified fluid.

\section{INTRODUCTION} \label{sec:introduction}

The motion of a submarine through a stratified ocean produces an
internal wave field that could be used, if accurately forecasted, for
detection and tracking.  These waves are produced by the vertical
displacement of the fluid as it flows over the submarine body and by the
disturbance of the fluid by the motions in the submarine's wake.  The
wake motions consist of turbulent eddies and the bulk motion due to the
collapse of the partially mixed wake region towards its equilibrium
density level.  In littoral regions of the ocean, where stratification
is strong and submarines travel slowly, the body-generated waves
predominate.  In the open ocean, where stratification generally is
weaker, the wake-turbulence generated waves are dominant.  

A very fast and accurate model has been developed to compute the
propagation of internal waves through realistic ocean environments. The
model, a modification of the methods described in \citet{BR04},
\citet{RBSD06} and \citet{RB08} involves a ray and caustic solution in
the Fourier transform domain, which is mapped into a spatial solution by
inverse Fourier transform.  This is a more practical approach than
calculating the ray and caustic solution directly in the spatial domain
and is general enough to treat background flows with depth dependent
shear and stratification.  

The propagation model requires initial conditions for the rays.  These
initial conditions can be provided by a near-field approximation of the
generation of the waves.  A convenient approximation is to represent the
waves emitted by a horizontally moving body as that due to a
horizontally moving and vertically oscillating distribution of sources
in a depth-dependent background.  The horizontally moving distribution
of sources models the body-generated waves and the vertical oscillation
of the source is an approximation of the unsteady wave generation by the
turbulent wake.  This idealization of the generation of the internal
wave field was proposed by \citet{Vo94} and \citet{DV96} and is based on
some ideas from the laboratory experiments of
\citeauthor{GT80} (\citeyear{GT81,GT82,GT83}).  Similar ideas have been
used by \citet{Ro97} for an ocean with a particularly sharp thermocline.
However, these theories for the parameterization of the near-field
internal wave field, and particularly the wave field generated by the
turbulent wake, have not been thoroughly validated.  

The focus of this paper is to use the results of numerical simulations to test
and revise the source distribution parameterizations of the near-field waves produced by submarines traveling in the littoral regions of the oceans.  The study is restricted to the idealized case of a sphere traveling horizontally at constant speed $U$ through a vertically stratified fluid at low Froude
number, \Fro, defined as $Fr = U/(Na)$ in which $a$ is a measure of the
radius of the body and $N$ is the (constant) buoyancy frequency of the
fluid.

The numerical simulations are for the stratified flow over and around
the body so that the internal wave field due to the displacement of the
fluid by the body is known.  The simulations are fully nonlinear, and
use a technique similar to the Cartesian-grid free-surface capturing
code (NFA) of \citet{DOWRWHYSAV07}.  The near wake solution is coupled
with the technique of \citet{DRIN02}, which efficiently simulates
turbulent wakes to distances in excess of 1000 body diameters down
stream, using the temporal approximation to simulate the internal waves
generated by the collapse of the turbulent wake.  Both the body
generated as well as the wake generated waves can be simulated.  This
allows a direct comparison between the numerically simulated wave field
and the near-field internal wave field computed from the parameterized
source distribution.

The Fourier-ray model, as specialized for the case of uniform stratification, is described in the next section, 
including the oscillating source distribution representation for flow produced by the
body. The numerical model is outlined in
following section,
along with a number of validation studies.  The comparison of the Fourier-ray model results with the
numerical simulations is described in the final section.

\section{THE FOURIER-RAY MODEL} \label{sec:raymodel}

The main aspects of the Fourier-ray model for horizontally homogeneous
background conditions that may have vertical variations of the buoyancy
frequency and currents are described in \citet{RBSM04a},
\citet{RBSM04b}, \citet{BRE03} and \citet{BR04}. Here our focus is on
the paramterization of the generation of linear internal waves by the
flow of a stratified fluid over and in the wake of a sphere. For this
purpose, we consider the simpler case of linear internal waves generated
by a body moving horizontally at a steady speed $U$ through a stably
stratified Boussinesq fluid with constant buoyancy frequency and no
background currents.   As a means of parameterizing the waves generated
by a turbulent wake, we will oscillate the sphere vertically at a
constant frequency $\sigma$, as discussed in more detail later in this
section.

The coordinate system is ${\bf  r} = (x,y,z)$, with $z$ positive
upwards, and is fixed to the mean position of the body. In this
reference frame, the background flow is ${\bf U} = (U, 0, 0)$. The
background buoyancy frequency is a constant value $N$. We solve for the
vertical displacement $\eta({\bf r},t)$. Any other dependent variable can
be computed from the vertical displacement using standard linear wave
polarization relations, as shown in \citet{Gi82}. 

An equation for $\eta({\bf r},t)$ can be derived from the linearized
Boussinesq equations of motion for a uniform background, as described
for example by \citet{Mi71},  \citet{Li78}, or \cite{BR04},
\begin{eqnarray}
  [ D^2  \partial_z^2  + (D^2 + N^2)
            (\partial_x^2 + \partial_y^2) ] \,
               \eta  = D \partial_z q .
          \label{Miles}
\end{eqnarray}
in which $D = \partial_t + U\partial_x$. The source distribution representing the body is specified in the form $q({\bf r},t) = q_0 ({\bf r})e^{-i \sigma t}$, in which $\sigma$ is the oscillation frequency of the source distribution.

The solution to this equation can be found by Fourier transform in the horizontal directions.
\begin{eqnarray}
 \eta ({\bf r},t) &=&  e^ {-i \sigma t} \,
                  \iint\limits_{ - \infty }^{~~~\infty}   
                       \tilde \eta (k,l,z) ~ e^{i (kx + ly) }  
          dk \, dl   ~.    
      \label{maslov}
\end{eqnarray}
where $ \tilde \eta (k,l,z)$ is the vertical eigenfunction,
\begin{eqnarray}
    \tilde \eta = {\rm sgn}(z) 2\pi i \hat \omega m Q_0 \, e^{imz}   /  B_m .  \label{Ieq2}
\end{eqnarray}
We are using subscript notation for partial derivatives, {\it e.g.} $B_m = \partial B/\partial m$, and $m$ is the internal wave vertical wavenumber given by the linear dispersion relation,
\begin{eqnarray}
      m =  \pm \, (k^2 + l^2)^{1/2} (N^2/ \hat\omega^2 - 1)^{1/2}  . \label{disprel}
\end{eqnarray} 
in which the internal wave wavenumber ${\bf k} = (k,l,m)$ and the intrinsic frequency $\hat \omega$ is given by
\begin{eqnarray}
  \hat \omega = \sigma - kU .
      \label{intfreq}
\end{eqnarray}  
The function B is, see \citet{Li78},
\begin{eqnarray}
    B = \hat\omega^2 m^2 - (N^2 - \hat\omega^2)(k^2 + l^2) .
  \label{Beq2}
\end{eqnarray}

Finally, $Q_0({\bf k})$ is the three-dimensional Fourier transform of $q_0$. Note that in $Q_0$, $m$ is treated as a function of $k,l$ through the dispersion relation.

As the factor $e^{-i\sigma t}$ accounts for all of the time-dependence in the present model, this solution can be considered as the long-time limit of an initial value problem in which the motion is started from rest and the body asymptotically in time oscillates vertically with constant frequency $\sigma$.

\subsection{Reflections from upper and lower boundaries}

To account for wave reflections from upper and lower horizontal boundaries, if they exist, we follow the method of \citet{BRE03}, which we outline here. For simplicity, we restrict attention to a background that is depth independent.  Each time a ray returns to any fixed depth $z$ above the body in a channel of total depth $H$ after reflecting once from the top of the channel and once from the bottom,  the wave phase $m z$ has changed by an amount $m 2H$. To account for the effects of this and subsequent reflections, we multiply $\tilde \eta (k,l,z)$ in (\ref{Ieq2}) by the sum
\begin{eqnarray}
      S &=& \sum_{n=0}^{\infty} ~ e^{i n 2 mH} \\ 
         &=&  i e^{-imH} / 2 \sin (mH) .
  \label{Sdef}
\end{eqnarray}
This is a divergent sum, since the individual terms do not vanish as $n \rightarrow \infty$, however the sum can be evaluated in the sense of generalized functions (see Eq. (1.2.2) of \citet{Ha49}. In any case, the sum diverges when $mH = \pi$,  when the interference of the reflected waves is perfectly constructive.  We eliminate this divergence by adding a small damping factor in the form of an imaginary wavenumber for the vertical wavenumber, or by limiting $S$ to a finite number of terms. The number of terms can be chosen to represent the number of reflections at a given time of interest, for given $k,l$,  as determined by a group velocity calculation. We have experimented with both methods but have used only the second method for the results presented here.

\subsection{Source distribution}

We consider internal waves generated by the vertically oscillating sphere of radius $a$ in a uniform flow of speed $U$. In the limit of large Froude number $Fr = U/Na$, as shown by \citet{GT82} and \citet{DV96}, the flow associated with this motion can be represented by the following source distribution function
\begin{eqnarray}
   q ({\bf r}, t) = -\frac{3}{2}
         \left[ U \frac{x}{a} + h \sigma e^{-i \sigma t}
            \frac{z}{a} \right] \delta (r - a) ,
      \label{msource}
\end{eqnarray}
where $\delta$ is the Dirac delta function. The vertical displacement amplitude of the oscillation of the sphere is $h$, and its vertical velocity is $h \sigma e^{-i \sigma t}$. Its three-dimensional Fourier transform $Q$ is
\begin{eqnarray}
   Q ({\bf k}, t) = \frac{3}{4} i \pi^{-2} a^3
         \left[ U k + h \sigma e^{-i \sigma t} m
          \right] \frac{j_1(Ka)}{Ka} ,
      \label{msource2}
\end{eqnarray}
where $K = | {\bf k} |$, and $j_1(z) = (\textrm{sin} z)/z^2 - (\textrm{cos} z)/z$ is the spherical Bessel function of order unity. 

The non-oscillating portion of the source distribution is an accurate representation of non-stratified flow over a sphere, but has been shown (and we will show here) that it also is a very good representation of the flow over sphere even for moderate to low Froude numbers. The oscillating portion of the source distribution has been used to model internal wave generation by turbulent eddies in the wake of an obstacle (\citet{DV96} based on some experimental work by \citet{GB85}). Following the guidelines of \citet{DV96}, the dominant internal waves generated by the eddies in the turbulent wake are simulated by choosing a value of $0.2$ for the Strouhal number $St = a \sigma / pi U$  and a source frequency of $\sigma = N \pi St Fr$. The amplitude of the vertical oscillation of the source, represented by the factor $h$ in (\ref{msource}), is at this stage chosen to match the amplitude of the observed waves. 

\subsection{Numerical procedure}

Putting this altogether, the spatial solution is found at any specified height $z$ by evaluating the ray solution (\ref{Ieq2}),  with $B$ given by (\ref{Beq2}) and $Q$ given by (\ref{msource2}), and multiplying the result by (\ref{Sdef}), and finally taking the inverse Fourier transform.

\section{THE NUMERICAL MODEL} \label{sec:numericalmodel}

The computer code Numerical Flow Analysis (NFA), \citet{DOWRWHYSAV07},
originally designed to provide turnkey capabilities to simulate the
free-surface flow around ships, has been extended to have the ability to
perform high fidelity stratified sub-surface calculations. The governing
equations are formulated on a Cartesian grid thereby eliminating
complications associated with body-fitted grids. The sole geometric
input into NFA is a surface panelization of the body. No additional
gridding beyond what is used already in potential-flow methods and
hydrostatics calculations is required. The ease of input in combination
with a flow solver that is implemented using parallel-computing methods
permit the rapid turn around of numerical simulations of high-\Rey
\hspace{1pt} stratified fluid-structure interactions.

The grid is stretched along the Cartesian axes using one-dimensional
elliptic equations to improve resolution near the body. Away from the
body, where the flow is less complicated, the mesh is coarser.
Details of the grid stretching algorithm, which uses weight functions
that are specified in physical space, are provided in
\citet{KnuppSteinberg1993}.

\subsection{Governing Equations} 

Consider a turbulent flow in a stratified fluid.
Physical quantities are normalized by characteristic velocity ($U_o$), length ($L_o$),  time ($L_o/U_o$), density
($\Delta \rho$), and pressure ($\rho_o U_o^2$) scales, where $\rho_o$ is the reference density and $\Delta \rho$ is the change
in density over the characteristic length scale.  Let $\rho$ and $u_i$ respectively denote the normalized density and three-dimensional velocity field as a function
of normalized space ($x_i$) and normalized time ($t$).  
The conservation of mass is
\begin{equation} 
\frac{\partial \rho}{\partial t} +\frac{\partial u_j
\rho}{\partial x_j} = 0 \;\; . 
\label{eq:mass} 
\end{equation}
For incompressible flow with no diffusion,
\begin{equation} 
\frac{\partial \rho}{\partial t} +u_j \frac{\partial
\rho}{\partial x_j} = 0 \;\; .  
\label{eq:density1} 
\end{equation}
Subtracting \eqref{eq:density1} from \eqref{eq:mass} gives a solenoidal
condition for the velocity:
\begin{equation} 
\frac{\partial u_i}{\partial x_i} = 0 \;\; .
\label{eq:solenoidal} 
\end{equation}

For an infinite Reynolds number, viscous stresses are negligible, and the conservation of momentum is
\begin{multline} 
\frac{ \partial \rho u_i}{\partial t}+\frac{\partial}{\partial x_j}
\left(\rho u_j u_i \right)  =  \\
-\frac{\rho_o} {\Delta \rho} \left[ \frac{\partial p}{\partial x_i} + \rho Ri_B  \, \delta_{i3} +\tau_i \right]  \; , 
\label{eq:momentum} 
\end{multline}
where $p$ is the normalized pressure and $\tau_i$ is a normalized stress that will act tangential to the surface of the body.  $\delta_{ij}$
is the Kronecker delta function.   The sub-grid scale stresses are
modeled implicitly \cite{dommermuth08}.  $Ri_B$ is the bulk Richardson number defined as:
\begin{equation}
Ri_B \equiv \frac{\Delta \rho}{\rho_o}\frac{g L_o}{U_o^2} \, ,
\label{eq:RiB}
\end{equation}
where $g$ is the acceleration of gravity.  The bulk Richardson number is the ratio of buoyant to inertial forces. 

The normalized density is decomposed in terms of the constant 
reference density plus a small departure which is further split into a mean and a fluctuation:
\begin{equation}
\label{eq:density2}
\rho = \frac{\rho_o}{\Delta \rho}  + \overline{\rho} \left(x_3\right)+\tilde{\rho} \left(x_i,t\right)   .
\end{equation}
Here,  $\overline{\rho}\left(x_3\right)$ and $\tilde{\rho} \left(x_i,t\right)$ are respectively the normalized density stratification and the normalized density perturbation. 

The pressure, $p$ is then decomposed into the dynamic, $p_d$, and hydrostatic, $p_h$, components as:
\begin{equation}
\label{eq:pressure}
p= p_d + p_h \; .
\end{equation}
The hydrostatic pressure is defined in terms of the reference density and the density stratification as follows:
\begin{equation}
\label{eq:hydrostatic}
\frac{\partial p_h}{\partial x_i} = -( \frac{\rho_o}{\Delta \rho}  + \overline{\rho}) Ri_B \delta_{i3} \; .
\end{equation}

The substitution of \eqref{eq:density2} into \eqref{eq:mass} provides a new expression for the conservation of mass:
\begin{equation} 
\frac{\partial \tilde{\rho}}{\partial t} +\frac{\partial}{\partial x_j}  \left( u_j (\overline{\rho}+\tilde{\rho}) \right)= 0 \;\; . 
\label{eq:mass2} 
\end{equation}

The substitution of \eqref{eq:density2}-\eqref{eq:hydrostatic} into
\eqref{eq:momentum} and using \eqref{eq:mass} to simplify terms gives a
new expression for the conservation of momentum:  
\begin{eqnarray} 
\lefteqn{\frac{ \partial u_i}{\partial t}+\frac{\partial}{\partial x_j} \left( u_j u_i \right)  =}   \nonumber \\ 
& &  -\frac{1} {1+ \gamma (\overline{\rho}+\tilde{\rho})} \left[ \frac{\partial p_d}{\partial x_i} + \tilde{\rho} Ri_B  \, \delta_{i3} +\tau_i \right]  , 
\label{eq:momentum2} 
\end{eqnarray}
where $\gamma = \Delta \rho/\rho_0$ is the characteristic density difference divided by the reference density.   If $\gamma << 1$, a Boussinesq  approximation may be employed in the preceding equation to yield:
\begin{eqnarray} 
\frac{\partial u_i}{\partial t}+\frac{\partial}{\partial x_j} \left( u_j u_i \right)  = -\frac{\partial p_d}{\partial x_i} - \tilde{\rho} Ri_B  \, \delta_{i3} -\tau_i  \; .
\label{eq:momentum3} 
\end{eqnarray}
If the background stratification is linear, we let 
\begin{equation}
\Delta \rho = -L_o  \frac{d \overline{\rho}^d}{d x_3^d} \, ,
\label{eqn:change}
\end{equation}
where a supersript $d$ denotes a  dimensional variable.  For a linear background
stratification, $Ri_B=Fr_{o}^{-2}$, where
$Fr_o = U_o(N L_o)^{-1}$ is the internal Froude number and $N$ is the Brunt-V\"{a}is\"{a}l\"{a} frequency defined as:
\begin{equation}
\label{eqn:brunt}
 N^2 = -\frac{g}{\rho_o}  \frac{d \overline{\rho}^d}{d x_3^d} \, ,
\end{equation}
The bulk Froude number, a ratio of inertial to gravitational forces, is defined as:
\begin{equation}
Fr_b \equiv \frac{U_o}{\left(gL_o\right)^{1/2}}.
\label{eq:Fr}
\end{equation}
Multiplying $Ri_B$ and $Fr_b^2$, \eqref{eq:RiB} times \eqref{eq:Fr} squared, yields the ratio of buoyant to gravitational forces:
\begin{equation}
\label{gamma}
Ri_B{Fr}_o^2 = \frac{\Delta \rho}{\rho_0} = \gamma.
\end{equation}
When the background stratification is linear, $\gamma={Fr}_b^2/{Fr}_o^2$.

The momentum equations using either \eqref{eq:momentum2} or
\eqref{eq:momentum3}  and the mass conservation equation
\eqref{eq:mass2} are integrated with respect to time.    The divergence
of the momentum equations in combination with the solenoidal condition
\eqref{eq:solenoidal} provides a Poisson equation for the dynamic
pressure.  The dynamic pressure is used to project the velocity onto a
solenoidal field and to impose a no-flux condition on the surface of the
body.   The details of the time integration, the pressure projection, and the formulation of the
body boundary conditions, are described in the next three sections.

\subsection{Time Integration}
A second-order Runge-Kutta scheme is used to integrate with respect to
time the field equations for the velocity and density.  During the first
stage of the Runge-Kutta algorithm, a Poisson equation for the pressure
is solved:
\begin{eqnarray}
\label{eqn:poisson1}
\frac{\partial}{\partial x_i} \left[ \frac{1} {1+ \gamma (\overline{\rho}+\tilde{\rho}^k)} \right] \frac{\partial p_d^k}{\partial x_i} =\frac{\partial}{\partial x_i} \left(
\frac{u^k_i}{\Delta t}+R^k_i \right) \; ,
\end{eqnarray}
where $R^k_i$ denotes the nonlinear convective, buoyancy, and stress
terms in the momentum equation, \eqref{eq:momentum2}, at time step $k$.
$u^k_i$, $\tilde{\rho}^k$, and $p_d^k$ are respectively the velocity components, density, and dynamic pressure at time step $k$.  $\Delta t$ is the time step.
For the next step, this pressure is used to project the velocity onto a solenoidal field. The first prediction for the velocity field ($u^*_i$) is
\begin{eqnarray}
\label{eqn:runge1}
u^*_i=u^k_i+\Delta t \left( R^k_i- \left[ \frac{1} {1+ \gamma (\overline{\rho}+\tilde{\rho}^k)} \right] \frac{\partial p_d^k}{\partial x_i} \right) \; .
\end{eqnarray}
The density is advanced using the mass conservation equation \eqref{eq:mass2}:
\begin{eqnarray}
\tilde{\rho}^*= \tilde{\rho}^{k}- \Delta t \frac{\partial} {\partial
x_j} \left[ u^k_j (\overline{\rho}+\tilde{\rho}^k) \right]
\end{eqnarray}
A Poisson equation for the pressure is solved again during the second stage of the Runge-Kutta algorithm:
\begin{multline}
\label{eqn:poisson2}
\frac{\partial}{\partial x_i} \left[ \frac{1} {1+ \gamma
(\overline{\rho}+\tilde{\rho}^*)} \right] \frac{\partial p_d^*}{\partial
x_i} = \\
\frac{\partial}{\partial x_i} \left(
\frac{u^*_i+u^k_i}{\Delta t}+R^*_i \right) \; ,
\end{multline}
$u_i$ is advanced to the next step to complete one cycle of the Runge-Kutta algorithm:
\begin{eqnarray}
\label{eqn:runge2}
\lefteqn{ u^{k+1}_i=\frac{1}{2} \left( u^*_i + u^k_i \right) }  \nonumber \\
& & +\frac{\Delta t}{2} \left( R^*_i -\left[ \frac{1} {1+ \gamma (\overline{\rho}+\tilde{\rho}^*)} \right] \frac{\partial p_d^*}{\partial x_i}  \right) \;\; ,
\end{eqnarray}
and the density is advanced to complete the algorithm:
\begin{eqnarray}
\tilde{\rho}^{k+1}= \frac{\tilde{\rho^*}+\tilde{\rho}^{k}}{2} -
\frac{\Delta t}{2} \frac{\partial} {\partial x_j} \left[ u^*_j
(\overline{\rho}+\tilde{\rho}^*) \right] \; .
\end{eqnarray}

\subsection{Enforcement of No-Flux Boundary Conditions}
  A no-flux condition is satisfied on the surface of the submerged body using a
finite-volume technique. 
\begin{equation}
\label{noflux}
u_i n_i = v_i n_i \; ,
\end{equation}
where $n_i$ denotes the unit normal to the body that points into the
into the fluid and $v_i$ is the velocity of the body. Cells near the body may have an irregular
shape, depending on how the surface of the body cuts the cell.   Let
$S_b$ denote the portion of the cell whose surface is on the body, and
let $S_o$ denote the other bounding surfaces of the cell that are not on
the body.   Gauss's theorem is applied to the volume integral of \eqref{eqn:poisson1}:
\begin{eqnarray}
\label{eqn:integral1}
\lefteqn{ \int_{S_o+S_b} ds   \left[ \frac{n_i} {1+ \gamma (\overline{\rho}+\tilde{\rho}^k)} \right]  \frac{\partial p_d^k}{\partial x_i}  =} \nonumber \\
& &  \int _{S_o+S_b} ds \left( \frac{u^k_i n_i}{\Delta t}+R_i n_i \right)  .
\end{eqnarray}
Here, $n_i$ denotes the components of the unit normal on the surfaces
that bound the cell.   Based on \eqref{eqn:runge1}, a Neumann condition is derived for the pressure on $S_b$ as follows:  
\begin{multline}
\label{eqn:bc1}
\left[ \frac{n_i}{1+ \gamma (\overline{\rho}+\tilde{\rho}^k)} \right]
\frac{\partial p_d^k}{\partial x_i} = \\
-\frac{u^*_i n_i}{\Delta t} +\frac{u^k_i n_i}{\Delta t}+R_i n_i \;\; .
\end{multline}
The Neumann condition for the velocity \eqref{noflux} is substituted into the preceding equation to complete the Neumann condition for the pressure on $S_b$:
\begin{multline}
\label{eqn:bc2}
\left[ \frac{n_i}{1+ \gamma (\overline{\rho}+\tilde{\rho}^k)} \right]
\frac{\partial p_d^k}{\partial x_i} =\\
-\frac{v^*_i n_i}{\Delta t} +\frac{u^k_i n_i}{\Delta t}+R_i n_i \;\; .
\end{multline}
This Neumann condition for the pressure is substituted into the integral formulation in \ref{eqn:integral1}:
\begin{eqnarray}
\label{eqn:integral2}
\lefteqn{ \int_{S_o} ds   \left[ \frac{n_i} {1+ \gamma (\overline{\rho}+\tilde{\rho}^k)} \right]  \frac{\partial p_d^k}{\partial x_i}  =} \nonumber \\
& &  \int _{S_o} ds \left( \frac{u^k_i n_i}{\Delta t}+R_i n_i \right) +  \int _{S_b} ds  \frac{v^*_i n_i}{\Delta t}  \; .
\end{eqnarray}
This equation is solved using the method of fractional areas.  Details
associated with the calculation of the area fractions are provided in
\citet{sussman01} along with additional references.  Cells whose cut
volume is less than 2\% of the full volume of the cell are merged with
neighbors.  The merging occurs along the direction of the normal to the
body. This improves the conditioning of the Poisson equation for
the pressure.   As a result, the stability of the  projection operator
for the velocity is also improved (see Equations \ref{eqn:runge1} and
\ref{eqn:runge2}). 

\subsection{Enforcement of No-Slip Boundary Conditions}

The stress $\tau_i$ is used to impose partial-slip and no-slip conditions on the surface of the body using a body-force formulation as follows:
\begin{equation}
\label{body}
\tau_i = \beta ( u_i -v_i) \delta ( {\bf x} - {\bf x_b^+} ) \; ,
\end{equation}
where $\beta$ is a body-force coefficient, $v_i$ is the velocity of the
body, ${\bf x}$ is a point in the fluid, and ${\bf x_b^+}$ is a point
slightly outside the body.  Equation~\eqref{body} forces the fluid
velocity to match the velocity of the body.  Note that free-slip
boundary conditions are recovered with  $\beta=0$ and no-slip boundary
conditions are imposed as  $\beta \to \infty$.   

\citet{dommermuth98} discuss modeling using body-force formulations.
Recently, there have been several studies which use a similar body
boundary conditions in both finite volume (\citet{Meyer:2010},
\citet{Meyer:2010a} and finite element (\citet{Hoffman:2006},
\citet{Hoffman:2006a}, \citet{John:2002} simulations.  The finite
element simulations of \citet{Hoffman:2006a} at very high Reynolds
numbers are able to predict the lift and drag to within a few percent of
the consensus values from experiments, using a very economical number of
grid points.  The finite volume implementations have also shown promise
albeit at more modest Reynolds numbers ($Re=3,900$).

In the following sections it is shown that finite volume simulations
with $Re \to \infty$ are able to accurately predict flow about
bluff-bodies using a partial-slip type boundary condition to model the
effects of the unresolved turbulent boundary layer.

\section{Validation} \label{sec:valid}

As discussed in \citet{dommermuth08} the advective terms in the momentum
equations are handled with the flux based limited QUICK scheme of
\citet{leonard97}.   Here, this treatment is extended to include the
advective terms in the density equation.  In this section three types of
numerical simulations are performed to assess the validity of
different aspects of the computational model proposed in the preceding
section.  First, a canonical turbulent shear layer is simulated to test
the ability of NFA to accurately resolve highly turbulent flows, while
remaining stable in the absence of explicit turbulence models, or
molecular viscosity.  Second, the unstratified flow over a sphere is
considered to test the ability of NFA to accurately predict flow
separation on a curved bluff body.  Lastly, the stratified flow over a
sphere is considered at several values of internal Froude number
ranging from $0.1-2.0$ to ensure that the effects of density
stratification are accurately captured.
%
%
\subsection{Turbulent Shear Layer}
\label{sec:TSL}
%
\begin{figure}
\begin{center}
\includegraphics[trim=0mm 0cm 0mm
0cm,clip=true,angle=0,width=\linewidth]{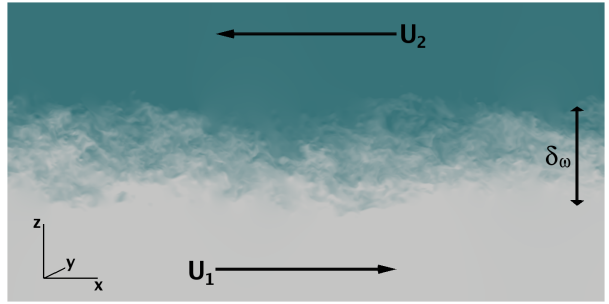}
\caption{NFA simulation of a turbulent mixing layer.  The illustration
is an $x-z$ plane of instantaneous streamwise, $x$, velocity.  Note the
wide range in the scales of motion, which is typical of shear generated
turbulence.}
\label{fig:2layer_still}
\end{center}
\end{figure}
  The first type of simulation used to validate NFA is a temporally
evolving shear layer.  An example of this type of flow is shown in
Figure~\ref{fig:2layer_still} to provide a visual reference to the
amount of disorder these flows generate, and a reference to the problem
setup. 

  A shear layer develops when two streams of fluid with different
velocities are brought together.  This type of flow is representative of
many engineering and geophysical flows.  The analysis of this type of
flow is simplified because the statistical description of the flow
reduces to a single dimension.  It is often considered a canonical flow
to study fundamental aspects of turbulence, and has been extensively
studied both experimentally (\citet{CorcosS:1976}, \citet{BellM:1990}
and \citet{SpencerJ:1971}, amongst others) and numerically
(\citet{RogersM:1994}, \citet{PantanoS:2002} and
\citet{BruckerS:2007},
again amongst others).  These detailed studies provide a rich database
to compare with, and these comparisons serve to assess the ability of
NFA to simulate highly turbulent flows.

  Here, the velocity streams, $U_1$ and $U_2$, are of equal magnitude
and opposite sign, with the velocity difference across the two streams
being $\Delta U \equiv U_1 - U_2$.  The model flow evolves temporally in
a domain that moves with the mean convective velocity $U_C = \left(U_1 +
U_2 \right)/2=0$. The thickness of the mixed region grows in time and
can be characterized by the vorticity thickness, $\delta_{\omega}(t)$ and
then momentum thickness, $\delta_{\theta}(t)$, respectively defined as:
\begin{equation}
\delta_{\omega}(t) \equiv \frac{\Delta U}{\left( d\left<u\right>/dx_3
\right)_{max}},
\label{eq:vthick}
\end{equation}
\begin{equation}
\delta_{\theta}(t) = \int^{\infty}_{-\infty} \left( \frac{1}{4}-  \left<
u_1\right>^2/\Delta U^2 \right) dx_3 .
\label{eq:mthick}
\end{equation}
In the preceding equations the $\left< \cdot \right>$ operator denotes
an average over the $x_1 - x_2$ plane, and fluctuations with respect to
the mean are obtained by applying a Reynolds decomposition, $u_i =
\left< u_i \right>  + u'_i$.

\subsubsection{Initial Conditions}
\label{sec:2layerICs}
\begin{figure}
\begin{center}
\includegraphics[width=\linewidth]{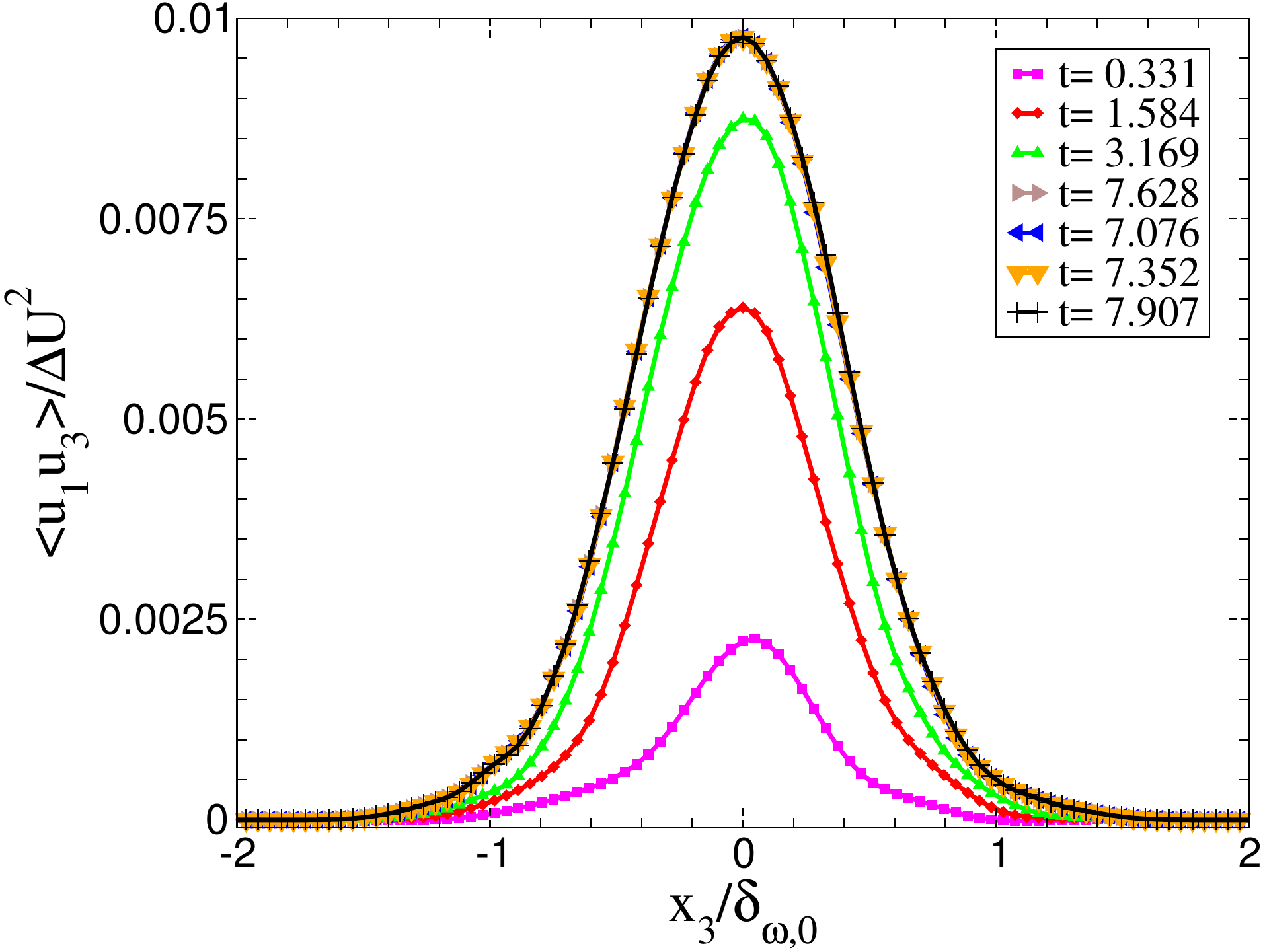}
\caption{Reynolds shear stress responsible for turbulent production at
several times during the adjustment phase.}
\label{fig:ics}
\end{center}
\end{figure}

  The initial conditions are constructed as a mean component plus a
small random disturbance.  The adjustment procedure of 
\citet{DRIN02}, in which the mean is held constant until the
production reaches a steady value, is used.  The adjustment time was $10
\delta_{\omega,0}/\Delta U$. Figure \ref{fig:ics} shows the $\left<u'_1 u'_3\right>$
cross-correlation responsible for turbulent production.  By $t=8$ the a
constant value was reached, and at $t=10$ the mean was allowed to
evolve.  This type of adjustment procedure removes the
arbitrary addition of ``turbulent like'' fluctuations, and allows
for repeatable initial conditions.

The mean initial conditions are:
\begin{multline}
\left< u_1 \right> = - \frac{\Delta U}{2} {\rm tanh} \left( \frac{2
x_3}{\delta_{\omega,0}} \right) ,\\
 \left< u_2 \right> = \left< u_3 \right> = \left< p \right> = 0.
\label{eq:means}
\end{multline}
where $\delta_{\omega,0}=1$ and $\Delta U = 1$.  

The amplitude of the random fluctuations added to the
mean velocity was $0.0165 \Delta U$.
%
  The computational domain $[L_1, L_2, L_3] =
[36.0, 12.0, 18.0]\delta_{\omega,0}$ was discritized with 
$[N_1, N_2, N_3]=[384, 128, 192]$ computational nodes.
The time step was $\Delta t = 0.001 \delta_{\omega,0}/\Delta U$.
Periodic boundary conditions were used in the stream-wise ($x,x_1$) and
span-wise ($y,x_2$) direction for all variables.  In the cross-stream
direction ($z,x_3$), Dirichlet boundary conditions were used for the
vertical velocity and Neumann boundary conditions were used for the other
velocity components and pressure.
%
%
%
%
\begin{figure}[ht!]
(a)\includegraphics[width=0.94\linewidth]{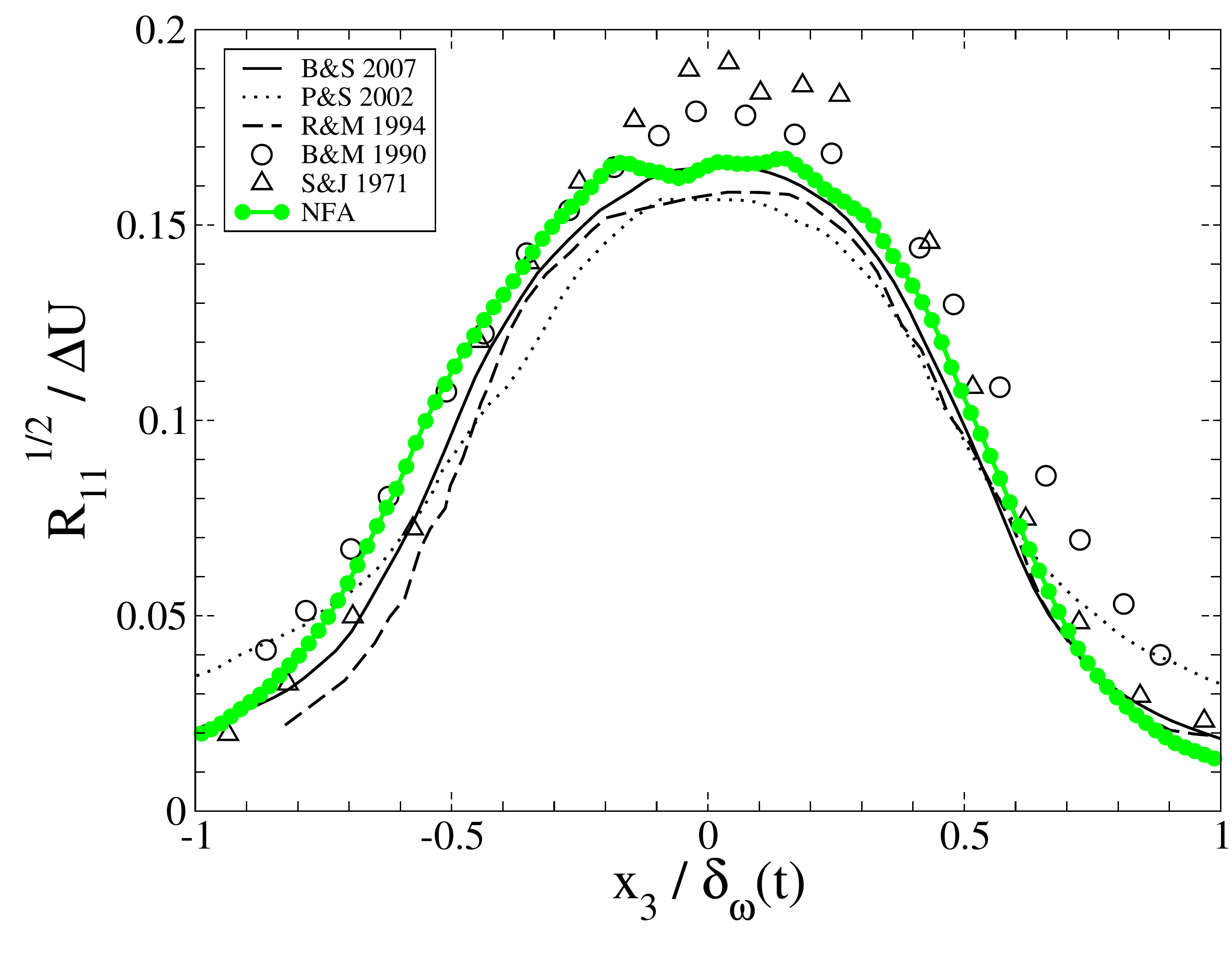}\\
(b)\includegraphics[width=0.94\linewidth]{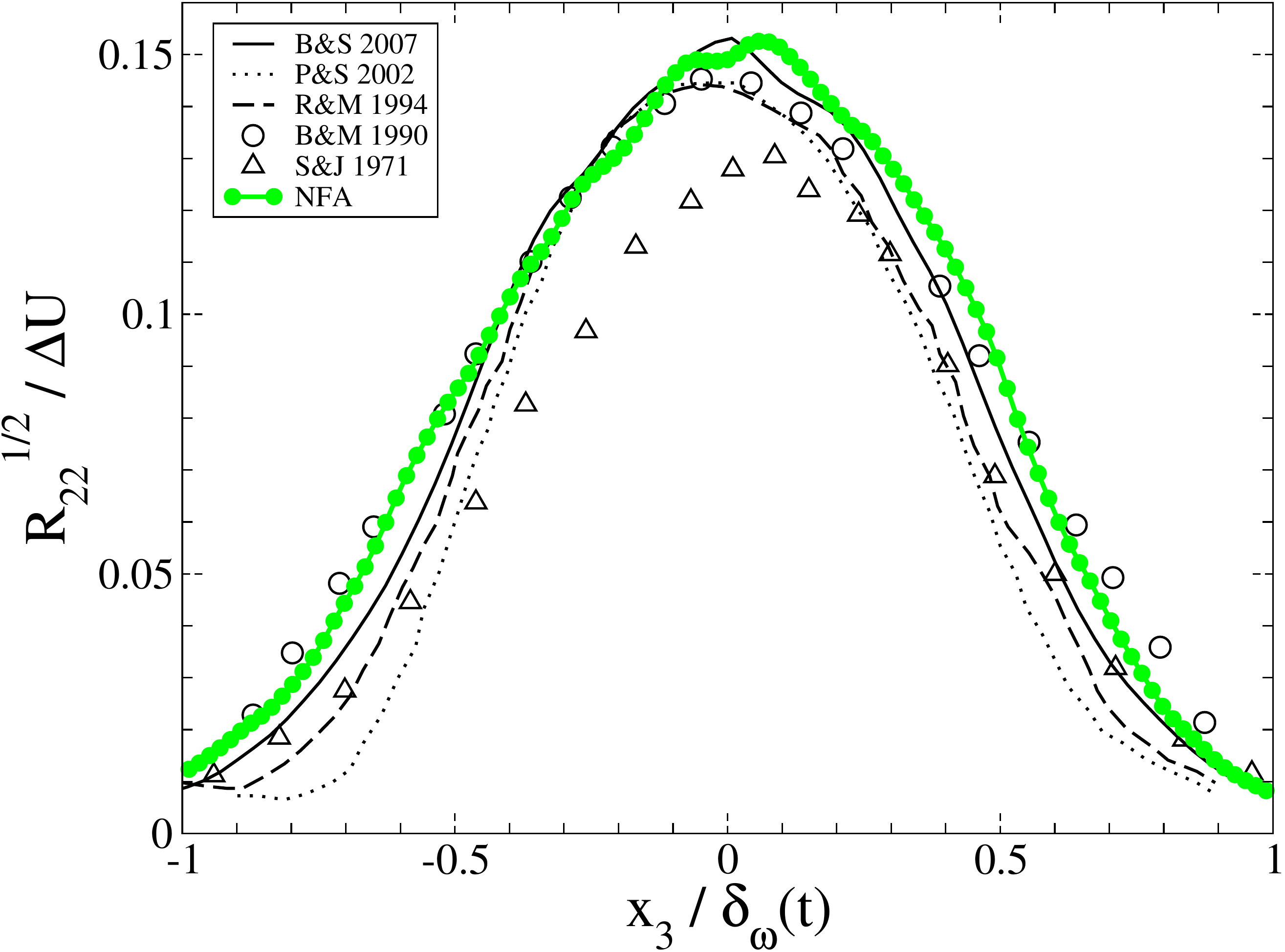}\\
(c)\includegraphics[width=0.94\linewidth]{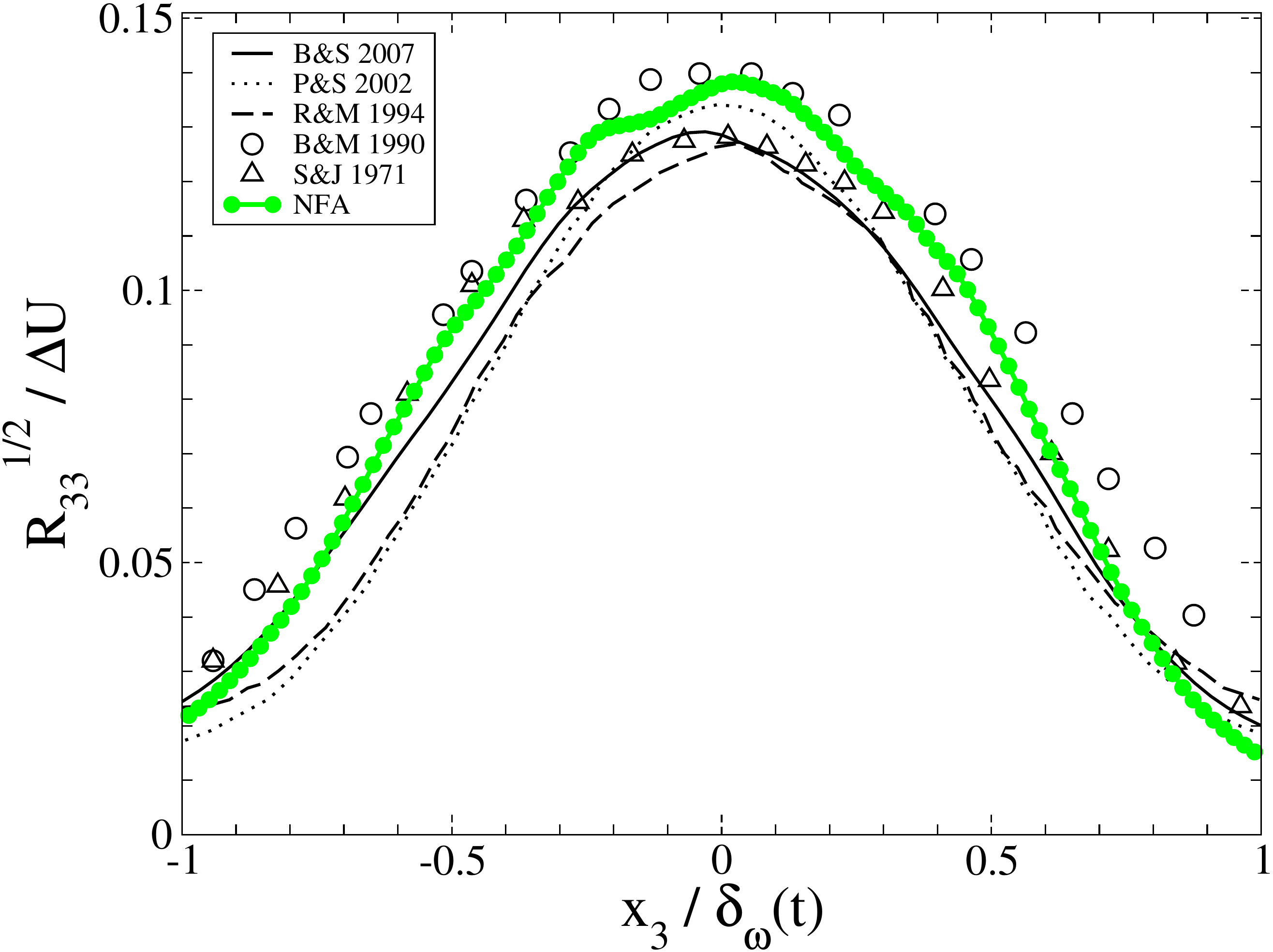}\\
\vspace{-10pt}
\caption{(a) Streamwise, (b)  Spanwise, (c) Cross-stream $r.m.s.$
velocities. The abbreviations in the legend
are as follows:  B\&S 2007 refers to the DNS of \protect \citet{BruckerS:2007},
P\&S 2002 refers to the DNS of \protect \citet{PantanoS:2002}, R\&M 1994 refers
to the DNS of \protect \citet{RogersM:1994}, B\&M 1990 refers to the
experiments of \protect \citet{BellM:1990}, and S\&J 1971 refers to the
experiments of \protect \citet{SpencerJ:1971}.} 
\label{fig:R11}
\end{figure}
%
\begin{figure}[ht!]
\begin{center}
(a)\includegraphics[width=0.94\linewidth]{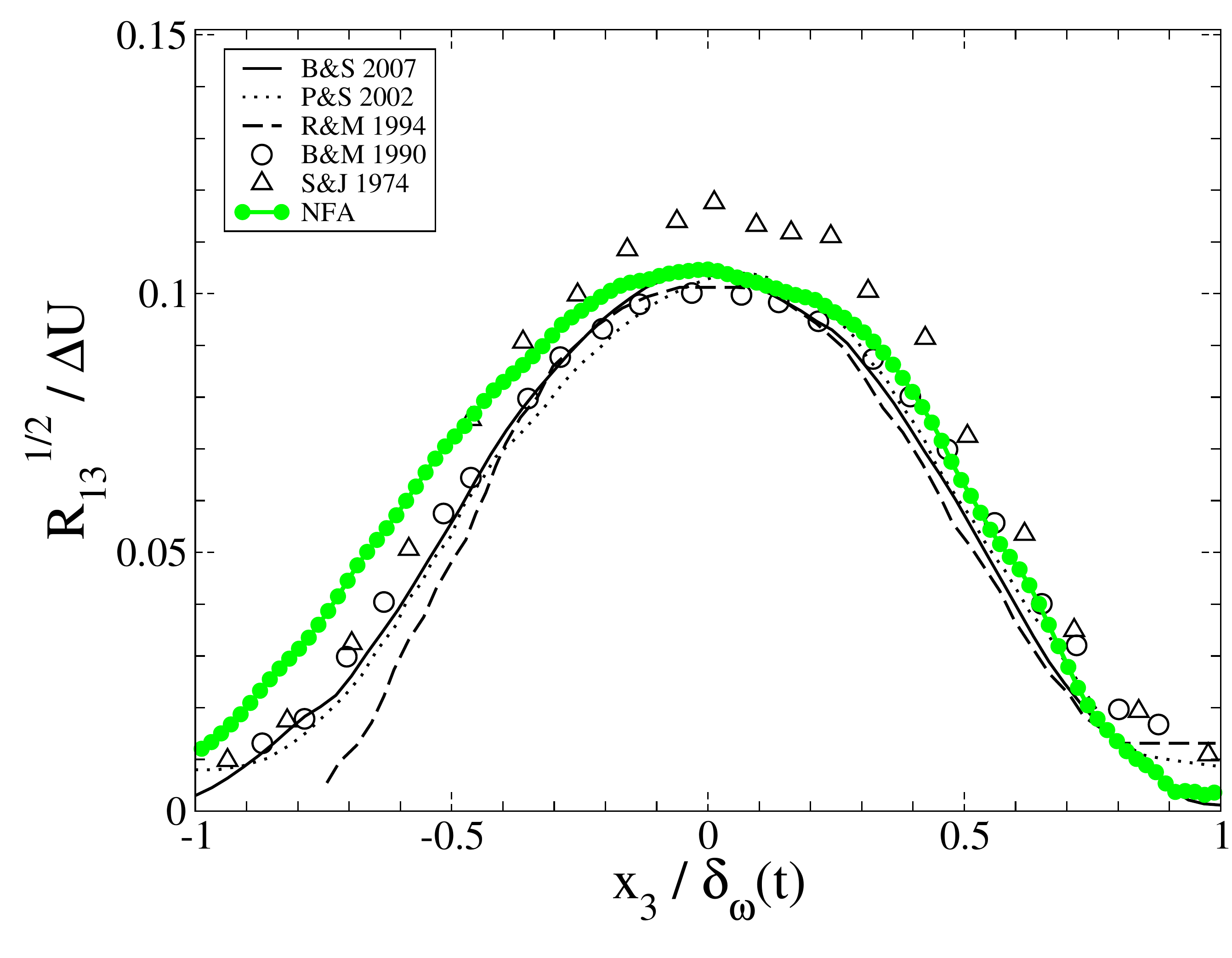}\\
(b)\includegraphics[width=0.94\linewidth]{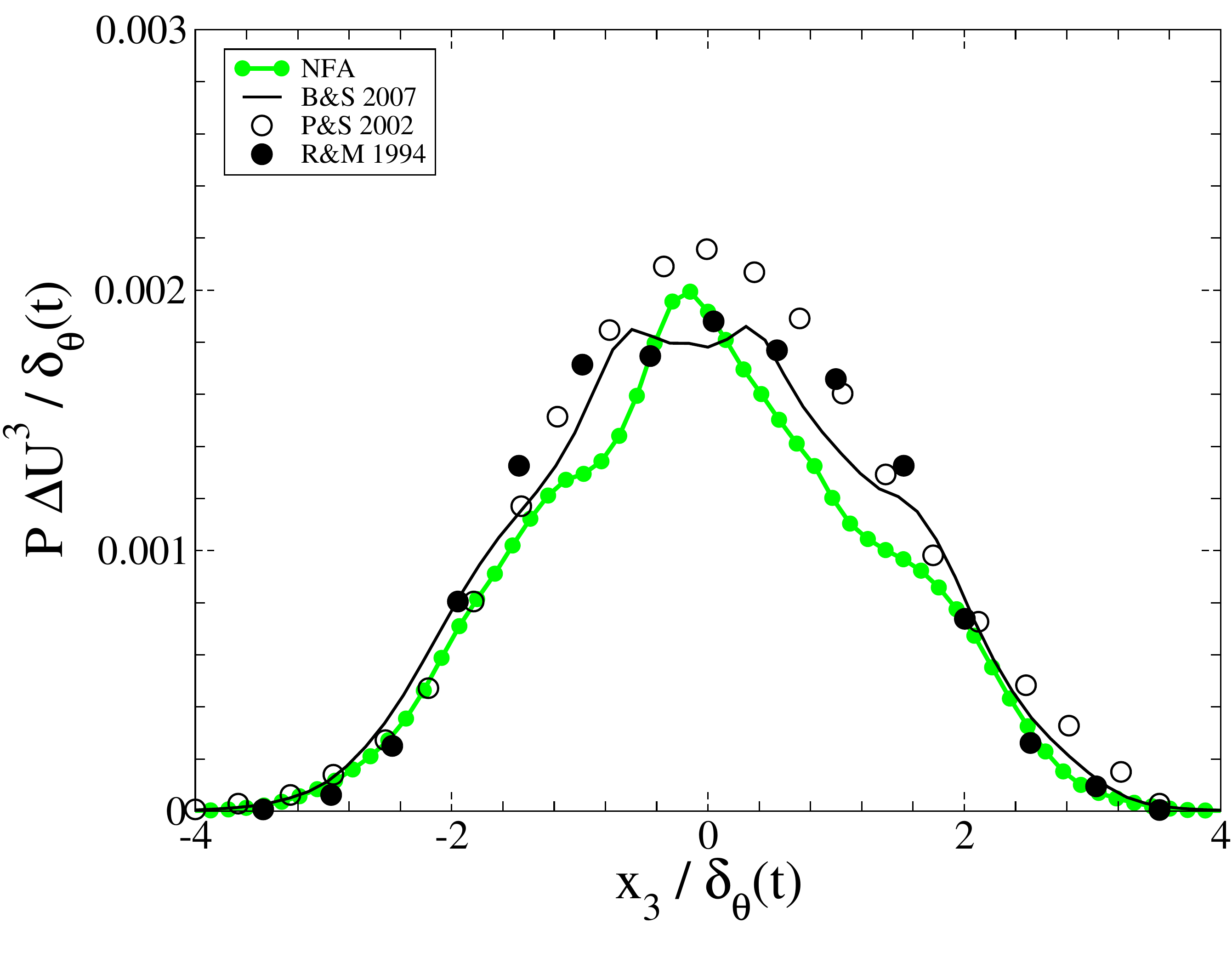}\\
\caption{(a) Reynolds shear stress $\left<u'_1 u'_3 \right>$. (b)
Turbulent production.  References in legend as in Figure
(\ref{fig:R11}).}
\label{fig:renstress}
\end{center}
\end{figure}

\subsection{Shear Layer Turbulence}
Figures~\ref{fig:R11}(a)-(c) compare the Reynolds
stresses, $R_{ij}=\left<u'_{i}u'_{j}\right>$ for $i=j$, obtained here with data
from previous studies. The Reynolds stresses were evaluated by averaging
the self-similar form at the following times: $\delta_{\omega,0}/U t=$ 110, 120, and 130
$\delta_{\omega,0} /\Delta U$. The peak streamwise, transverse, and
spanwise turbulent intensities, shown in
Figures~\ref{fig:R11}(a)-(c), $\sqrt{R_{11}}/\Delta
U = 0.17$, $\sqrt{R_{33}}/\Delta U = 0.14$, and $\sqrt{R_{22}}/\Delta U
= 0.15$ agree well with previous DNS and experimental data. The experimental data shows a
scatter of about $10 \%$ with respect to the simulation data. The shape
of the self-similar profiles also agrees well.
Comparisons of the the Reynolds stresses,
$R_{ij}=\left<u'_{i}u'_{j}\right>$ for  $i \ne j$, and the
 production of turbulent kinetic energy defined as: 
\begin{equation} 
P \equiv - \left< u'_{i} u'_{j} \right> \frac{\partial \left< u_{i}
\right> }{\partial x_{j}}  = - \left< u'_{1} u'_{3} \right>
\frac{\partial \left< u_{1} \right>} {\partial x_{3}} \; , 
\label{eq:P}
\end{equation}
in the current simulations to that from an incompressible
DNS (\citet{RogersM:1994},\citet{BruckerS:2007} and compressible low Mach number
DNS (\citet{PantanoS:2002} are shown in Figure~\ref{fig:renstress}(a)-(b),
respectively.  The results
from the current simulations agree well with previous DNS and laboratory data.  

In summary, the turbulent shear layer
simulations have been successfully bench marked against previous
laboratory and DNS data, and show NFA's ability to correctly capture the
energetically important scales of turbulent motion without requiring
explicit modeling. 

%
%
\section{Flow over a Sphere}
\begin{figure}
\begin{center}
\includegraphics[width=0.9\linewidth]{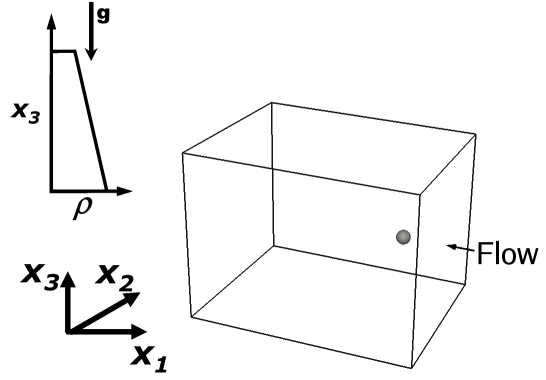}
\caption{A schematic of the computational domain.}
\label{fig:schematic}
\end{center}
\end{figure}

NFA simulations of flow over a sphere in uniform and stratified fluids
are respectively discussed.  The computational domain is shown
schematically in Figure~\ref{fig:schematic}. The size of the computational domain $[L_1, L_2, L_3]$, the number of
computational nodes $[N_1, N_2, N_3]$, the near body grid spacing,
$\Delta x_i^{NB}$, the time step, $\Delta t$, the internal Froude number,
$Fr$, and the near-body parameter $\beta$ are
provided in Table~\ref{tbl:sphere_params} for all simulations
subsequently discussed.

  All of the simulations used In-flow and Out-flow boundary conditions
in the stream-wise ($x_1$) direction, and Neumann boundary conditions in the
span-wise ($x_2$) direction.  In the cross-stream direction ($x_3$),
Dirichlet boundary conditions are used for the vertical velocity, $u_3$,
and Neumann boundary conditions are used for the other velocity components,
pressure, and density.  

 In the NFA simulations the flow is accelerated from rest by multiplying
the free-stream current with the following function:
\begin{equation}
f(t)=\left\{1-exp\left[-\left(t/T_{cur}\right)^2\right]\right\}.
\label{eq:ramp}
\end{equation} 
$T_{cur}=0.5$ was used in all simulations.

%
%
%
%
\begin{table*}[ht!] 
\begin{center}
\caption{\label{tbl:sphere_params} Parameters for all sphere simulations.  Domain size, grid points, grid spacing near the body,
time step, near body parameter, $\beta$.  All normalized with the body
diameter $D$ and free-stream velocity $U$.  The internal Froude number,
$Fr$, for each case is given by the case name. Cases labeled $FrX$ denote a domain
used for runs at different Froude numbers.}
\vspace{5pt}
\begin{tabular}{c|ccccccccc}
\hline
\hline
Case          & $L_x$ & $L_y$ & $L_z$ & $N_x$ & $N_y$ & $N_z$ & $\Delta t$ & $\beta$ & $\Delta x_i^{NB}$ \\
\hline
$Fr1_1$       &  19   &  20   &  10   &  512  & 384   & 384   & 0.0025     & 0.67    &  0.03        \\ 
$Fr1_2$       &  19   &  20   &  5    &  512  & 384   & 256   & 0.0020     & 0.33    &  0.03        \\  
$Fr4_1$       &  19   &  20   &  10   &  1024 & 768   & 768   & 0.00125    & 1.00    &  0.01        \\                  
$Fr4_2$       &  19   &  20   &  5    &  512  & 384   & 256   & 0.002      & 0.33    &  0.03        \\                    
$Fr{\infty}_1$&  10   &  8    &  8    &  1024 & 768   & 768   & 0.001      & 1.00    &  0.008       \\                
$Fr{\infty}_2$&  10   &  8    &  8    &  1024 & 768   & 768   & 0.001      & 0.00    &  0.008       \\                  
$FrX^C$       &  12   &  6    &  6    &  384  & 256   & 256   & 0.005      & 0.67    &  0.03        \\                  
$FrX^M$       &  12   &  12   &  12   &  384  & 384   & 384   & 0.005      & 0.67    &  0.03        \\                 
\hline 
\hline
\end{tabular}
\end{center}
\end{table*}

\subsection{Uniform fluid}
\label{sec:UnstratSphere}
\begin{figure*}
\begin{center}
\includegraphics[trim=0mm 0cm 0mm
0cm,clip=true,angle=0,width=\linewidth]{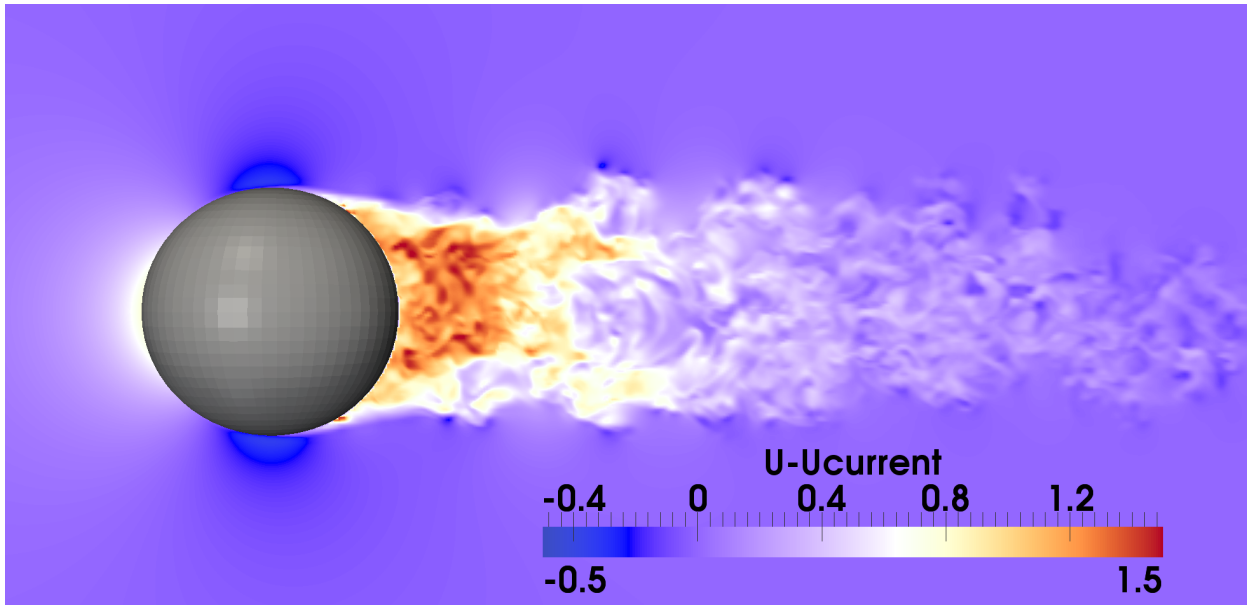}\\
\vspace{25pt}
\includegraphics[trim=0mm 0cm 0mm
0cm,clip=true,angle=0,width=\linewidth]{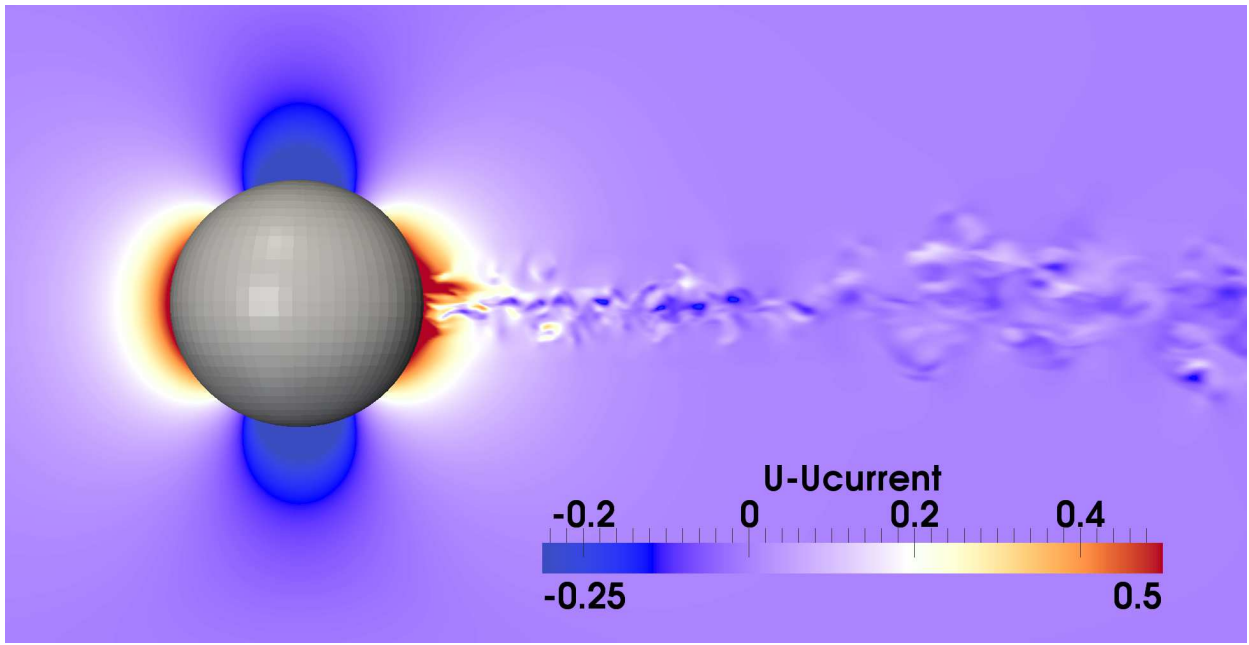}\\
\caption{NFA simulations of unstratified flow over a sphere. The
illustrations are of the instantaneous stream-wise velocity on the $x -
z$ plane located at $y=0$. Top case $Fr\infty_1$ with $\beta=1$.
Bottom case $Fr\infty_2$ with $\beta=0$.}
\label{fig:sphere_viz}
\end{center}
\end{figure*}

Contours of the instantaneous streamwise velocity on the $x-z$ plane
located at $y=0$ are shown in Figure~\ref{fig:sphere_viz} for cases $Fr\infty_1$ with
$\beta=1$ (top) and $Fr\infty_2$ with $\beta=0$ (bottom).  When
$\beta=0$ the boundary conditions on the body are no-slip and no flow
separation occurs.  When $\beta=1$ the flow separates just shy
$120\deg$, the value reported in \citet{Achenbach:1972}.  
  Here, a comparison of the drag coefficient to theory and
experiments is used to assess the quality of the flow near the body.
Since in the unstratified case the drag is determined by the fields in
the immediate vicinity of the body it is an excellent metric for
determing the quality of those fields.

The drag coefficient, is defined as:
\begin{equation}
C_D \equiv \frac{F_D}{1/2 \rho_0 U^2 A_P} \; ,
\label{eq:CD}
\end{equation}
where $F_D$ is the drag force and $A_P$ is the project frontal area of
the body.

   The drag force on the body is calculated by integrating the normal
pressure over the surface. \citet{Achenbach:1972} reports 
that the viscous contribution to the drag is $\sim 2\%$ of the total at
$Re= 5\times 10^6$, and hence the the total drag should be well approximated by
the pressure drag at high-\Rey.  The viscous contribution is not calculated. 

%
\begin{figure}
\begin{center}
\includegraphics[width=\linewidth]{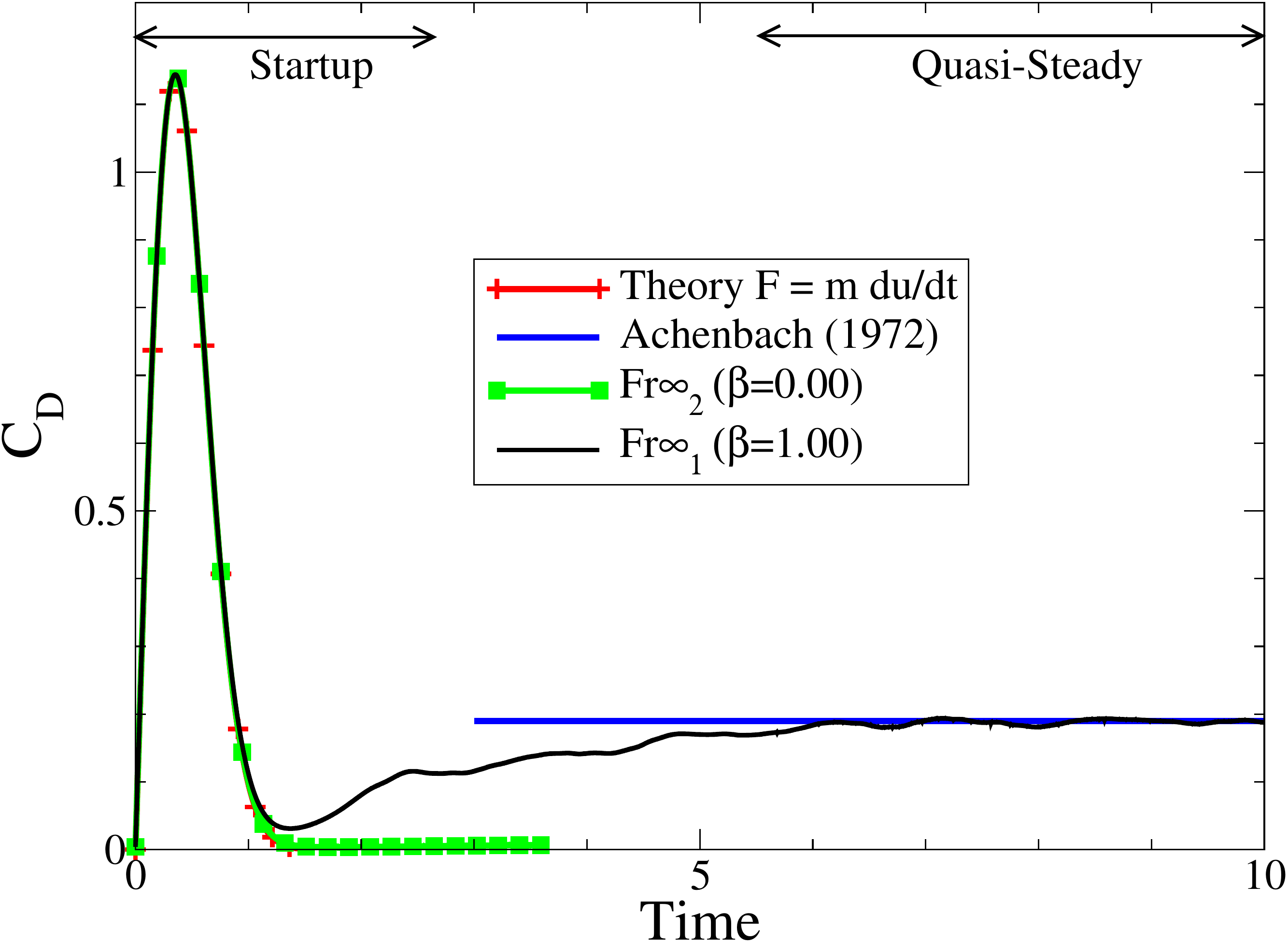}
\caption{Drag coefficient for flow over a sphere at high-Reynolds
number.  The references in the legend are: Achenbach (1972) refers to the
experiments of \protect \citet{Achenbach:1972} while the theoretical prediction
refers to equation~\ref{eq:added_mass}.} 
\label{fig:Cd_unstrat}
\end{center}
\end{figure}

  The coefficient of drag for cases $Fr\infty_1$ with $\beta=1$ and
$Fr\infty_2$ with $\beta=0$ are shown in Figure~\ref{fig:Cd_unstrat}.  
As noted in the discussion of Figure~\ref{fig:sphere_viz} the free-slip boundary conditions, 
used in case $Fr\infty_2$, did not allow for flow separation and hence
there was almost complete pressure recovery in the lee of the sphere and
hence no drag.  For case $Fr\infty_1$, in which separation occured at
the correct location, the drag coefficient was $0.2$ the same value reported
in the experiments of \citet{Achenbach:1972}. 

  During the startup phase of the simulations, $f(t)$ is non-zero and there is an added mass component of the drag:
\begin{equation}
F_D^{AM} = m U\frac{df}{dt}
\label{eq:added_mass}
\end{equation} 

  The drag due to the added mass is also shown in
Figure~\ref{fig:Cd_unstrat}, and the agreement between both simulations
and the prediction is excellent.

%
%
\subsection{Flow over a sphere in a stratified fluid} \label{sec:StratSphere}
  Having validated the drag on a sphere in the case without
stratification, attention is now turned to the case with stratification.
To proceed comparisons are made between existing theoretical and
experimental data sets. The stratification of the fluid can cause body
generated waves and significant drag at low Froude numbers ($\leq 1.5$),
while at moderate Froude numbers ($1.5\geq Fr \leq 3$)  the potential
energy gained overcoming the poles of the sphere aids in pressure
recovery in the lee and a small drag reduction.
%
\begin{figure}
\begin{center}
\includegraphics[width=\linewidth]{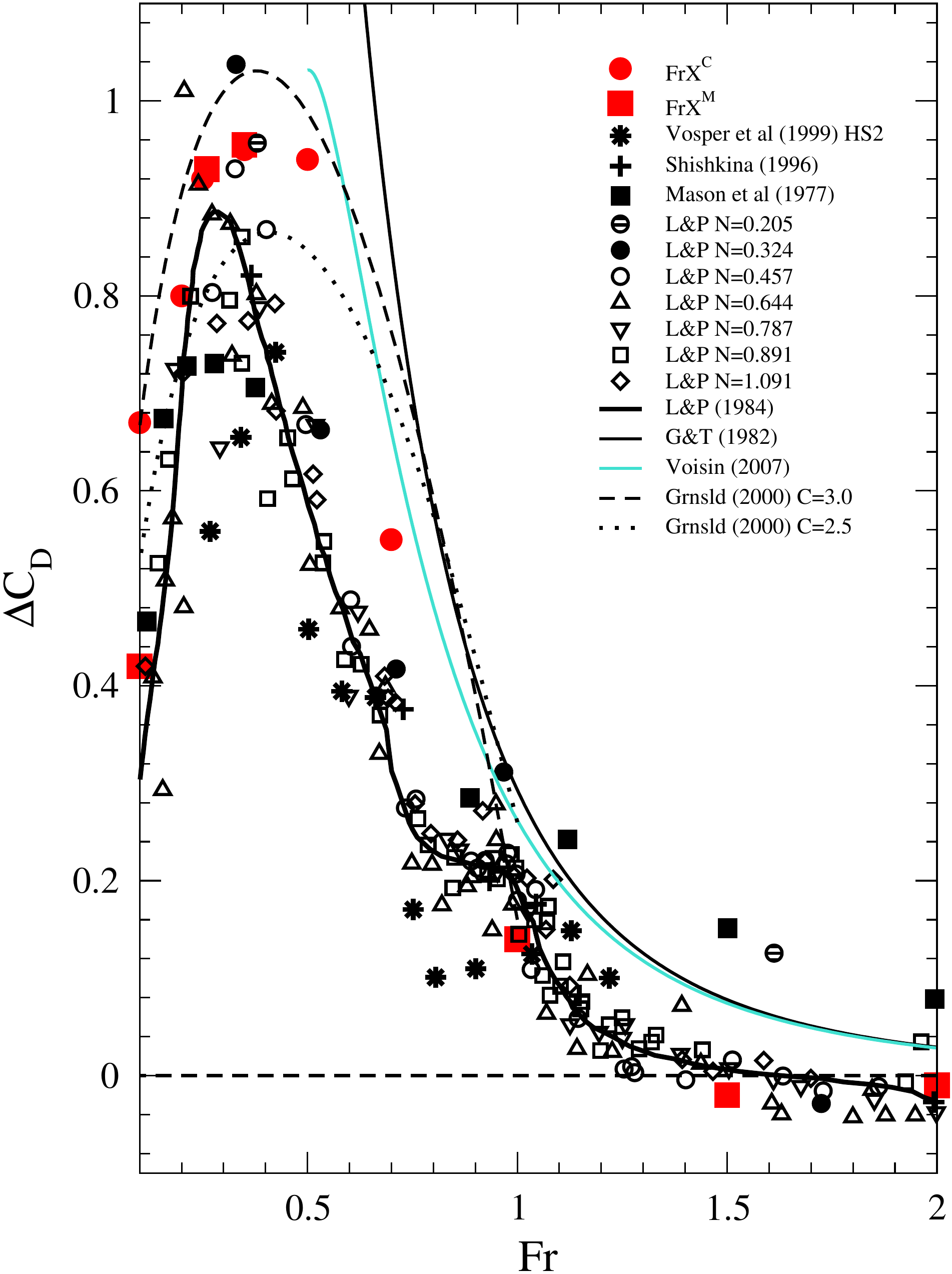}
\caption{Change in the drag coefficient as a function of internal Froude
number.  In the legend, $FrX^C$ and $FrX^M$ denote cases with the coarse
and medium resolutions reported in Table~\ref{tbl:sphere_params}.}
\label{fig:DeltaCd_Fr}
\end{center}
\end{figure}
Following, \citet{Lofquist:1984} the change in drag due
to stratification, $\Delta C_D$, is defined as:
\begin{equation}
\Delta C_D(Fr) = C_D(Fr) - C_D\left( Fr\infty \right).
\label{eq:CDFR}
\end{equation} 

Figure~\ref{fig:DeltaCd_Fr} shows the change in the drag
coefficient, $\Delta C_D$, as a function of the internal Froude number,
$Fr=U/(2r N)$. NFA simulations at
$Fr=0.1,0.2,0.25,0.35,0.5,0.75,1.0,1.5,2.0$ are compared with
the theories of \citet{Greenslade:2000}, \citet{Voisin:2007}, and
\citet{GT:1982} along with the experiments of \citet{Lofquist:1984},
\citet{Vosper:1999}, \citet{Shishkina:1996}, and \citet{Mason:1977}.

In summary, Figure~\ref{fig:DeltaCd_Fr} shows that NFA is able to predict correctly the change in drag for both low and moderate Froude numbers, and appears to correctly predict the transition from low to high Froude number regimes, which occurs between $Fr\approx0.7$ and $Fr\approx1.0$, although this prediction currently is based on only one computed data point.

\section{INTERNAL WAVES} \label{sec:internalwaves}
\begin{figure}
   \begin{center}
     \includegraphics[width=0.9\linewidth]{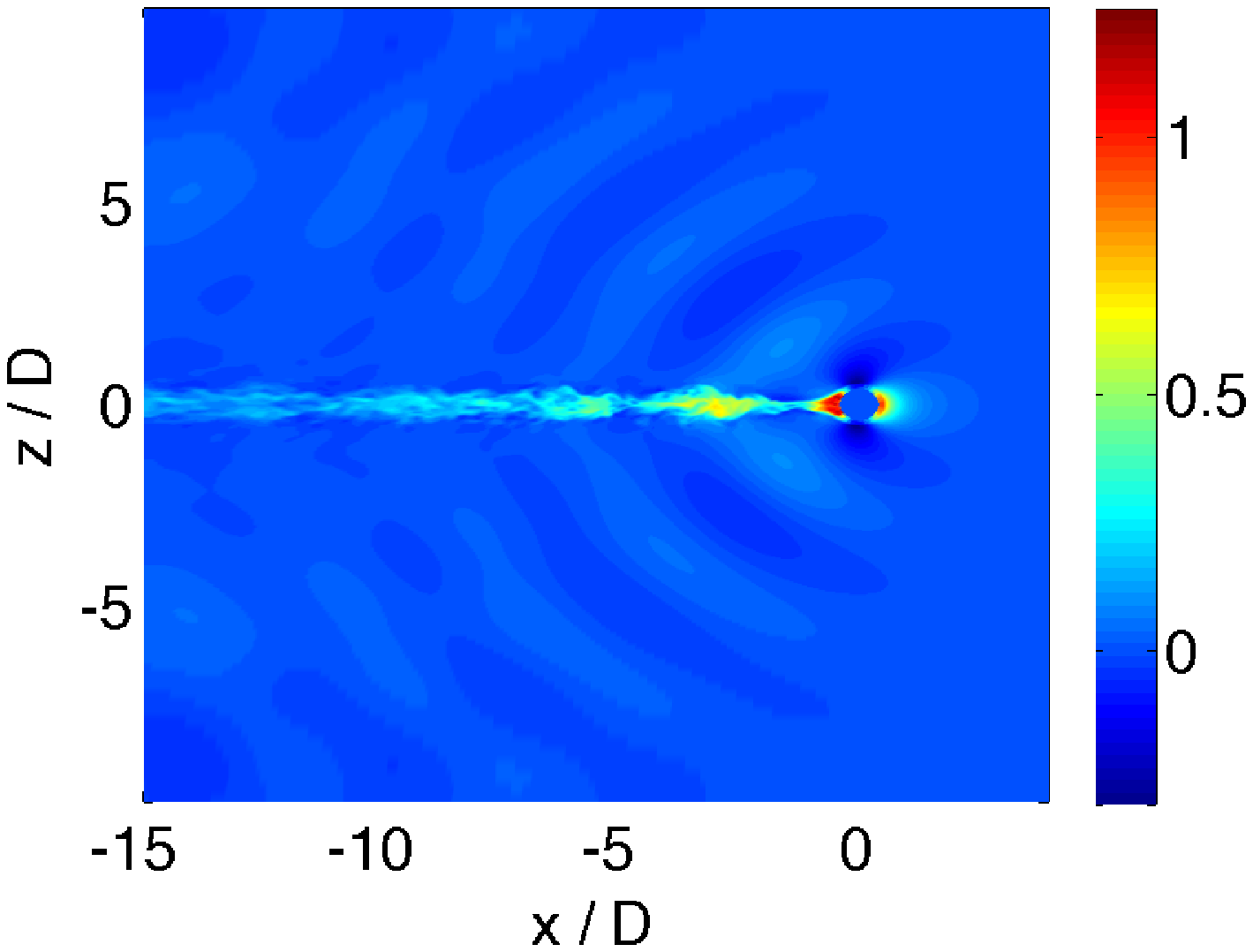}
     \includegraphics[width=0.9\linewidth]{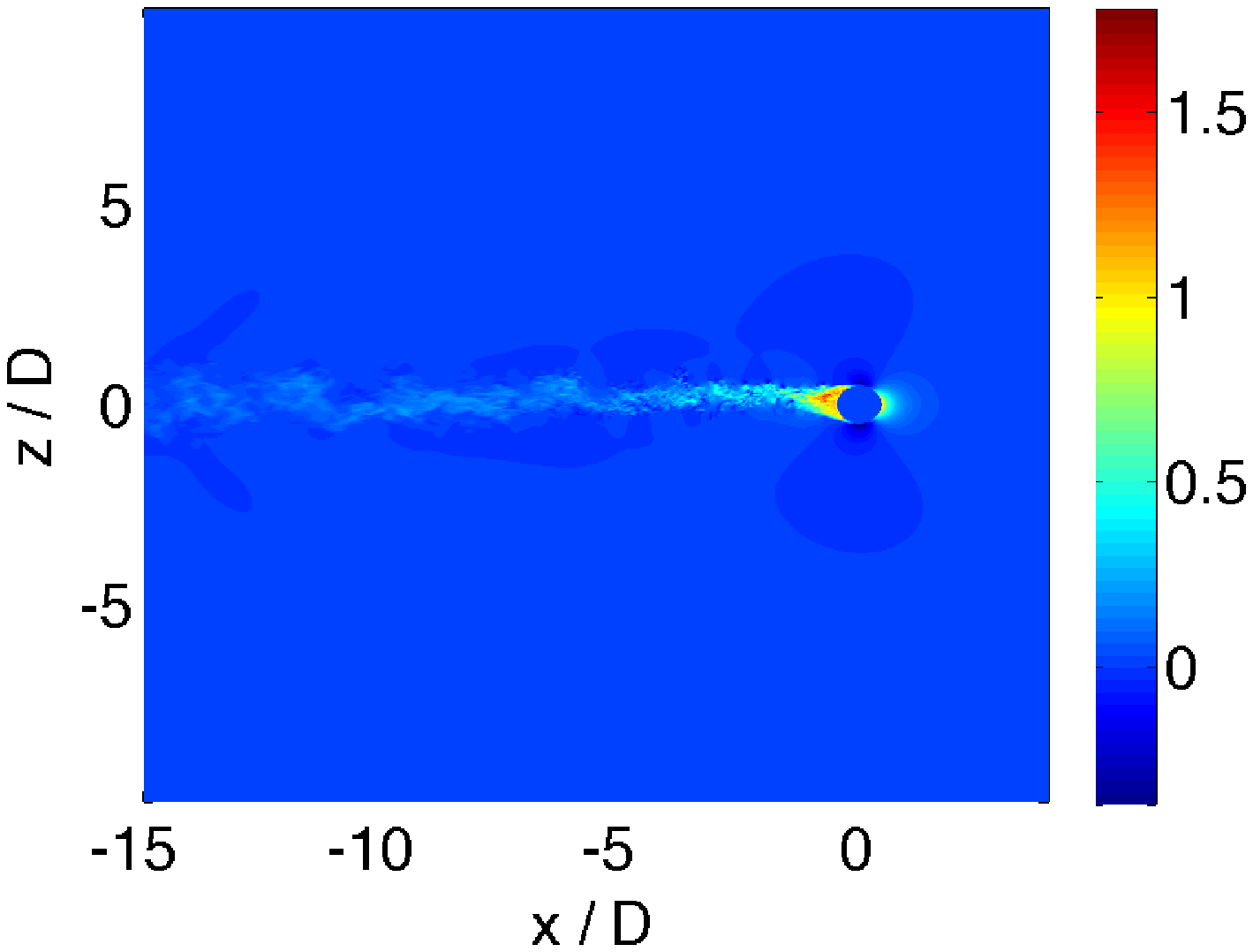}
   \end{center}
   \caption{Along track vertical $(x,z)$ plane at $y = 0$ of the numerical solutions at the wake center for $Fr = 1$ (upper panel) and $Fr = 4$ (lower panel).  The plotted variable is $u/U$, the $x$-component of the velocity of the wavefield.}
   \label{fr14}
\end{figure}

We now compare the internal waves computed by NFA for stratified flow
over a sphere with the field  predicted by linear wave theory, using an
oscillating source distribution to parameterize the generation of the
waves by the sphere and its wake, as discussed in The Fourier-Ray Model
section. As this is a preliminary study, we consider
flows for just two moderate Froude numbers, $Fr = 1$ and $4$. The sphere
is towed in the positive $x$ direction, and each plot shown below is
either a horizontal $(x,y)$ cross section at a height of two sphere
diameters ($z = 2D$) above the centerplane of the sphere or a vertical
$(x,z)$ plane through the wake centerline. The plots show the $x$
component $u$ of the wavefield velocity. 

For the Fourier-ray calculations, the inverse Fourier transform
(\ref{maslov}) was approximated discretely on a wavenumber grid of 1024
by 512 in $k,l$, for $Fr = 1$ and $Fr = 4$. In addition to resolving the
flow features, the extent of the wavenumber grid must be chosen to limit
periodic wrap-around errors, which result from the discrete
approximation of the inverse Fourier transform. For a single depth, the
theoretical solution takes only a few seconds to calculate on a standard
PC.

Figure~\ref{fr14} shows the horizontal velocity $u/U$ on a vertical
plane through the wake centerline ($y = 0$) for the two cases $Fr = 1$
(top pane) and $Fr = 4$ (bottom pane). The lower Froude number case
shows a well-defined lee wave field as well as a weakly turbulent wake.
The lee wave field is steady with respect to the sphere, but downstream
of about $x/D = -5$ internal waves generated by the turbulent wake can
be seen. For the higher Froude number, the lee waves are very weak and
the turbulent wake is stronger and shows evidence of periodic
fluctuations.

\begin{figure}
   \begin{center}
     \includegraphics[width=\linewidth]{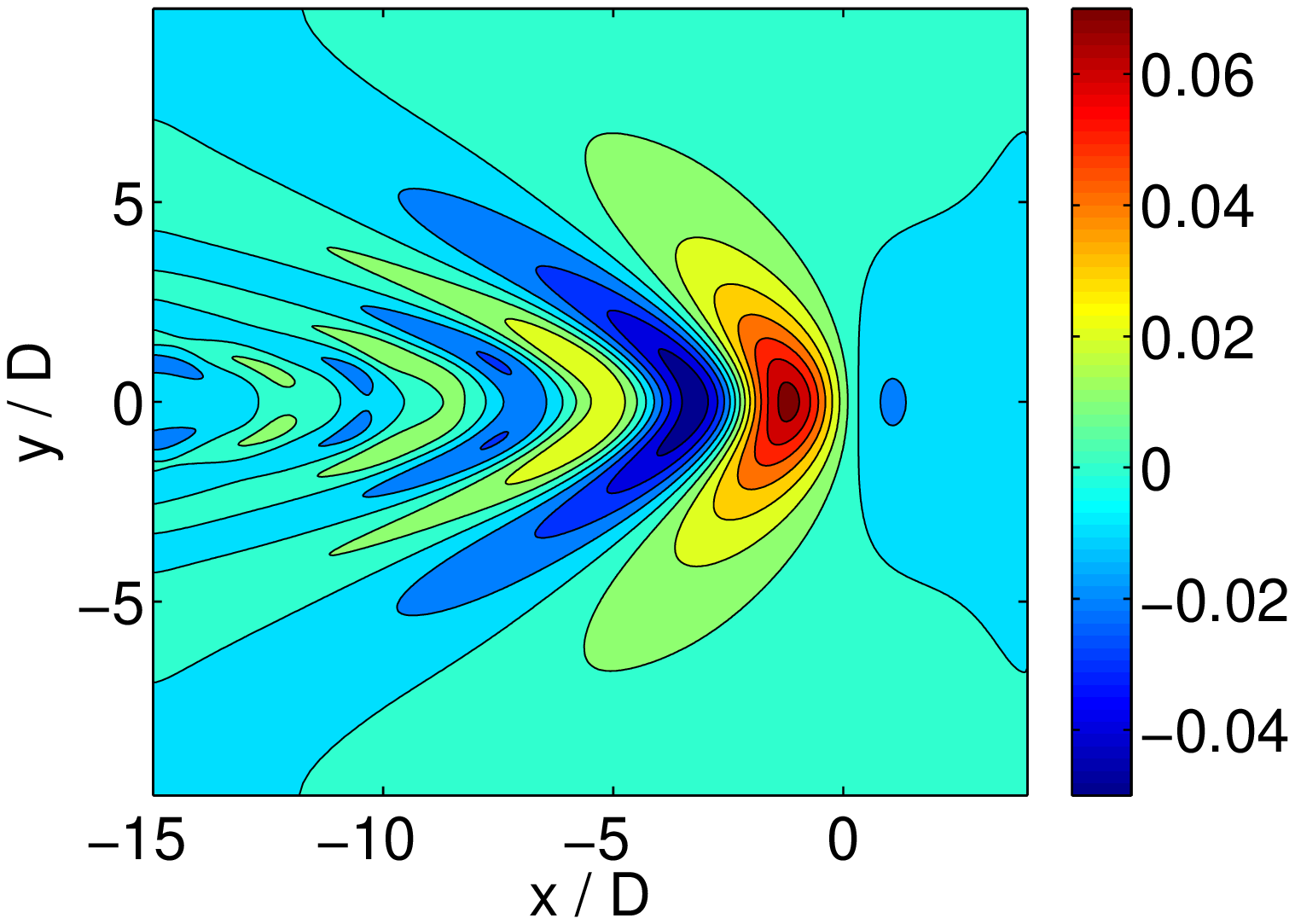}
     \includegraphics[width=\linewidth]{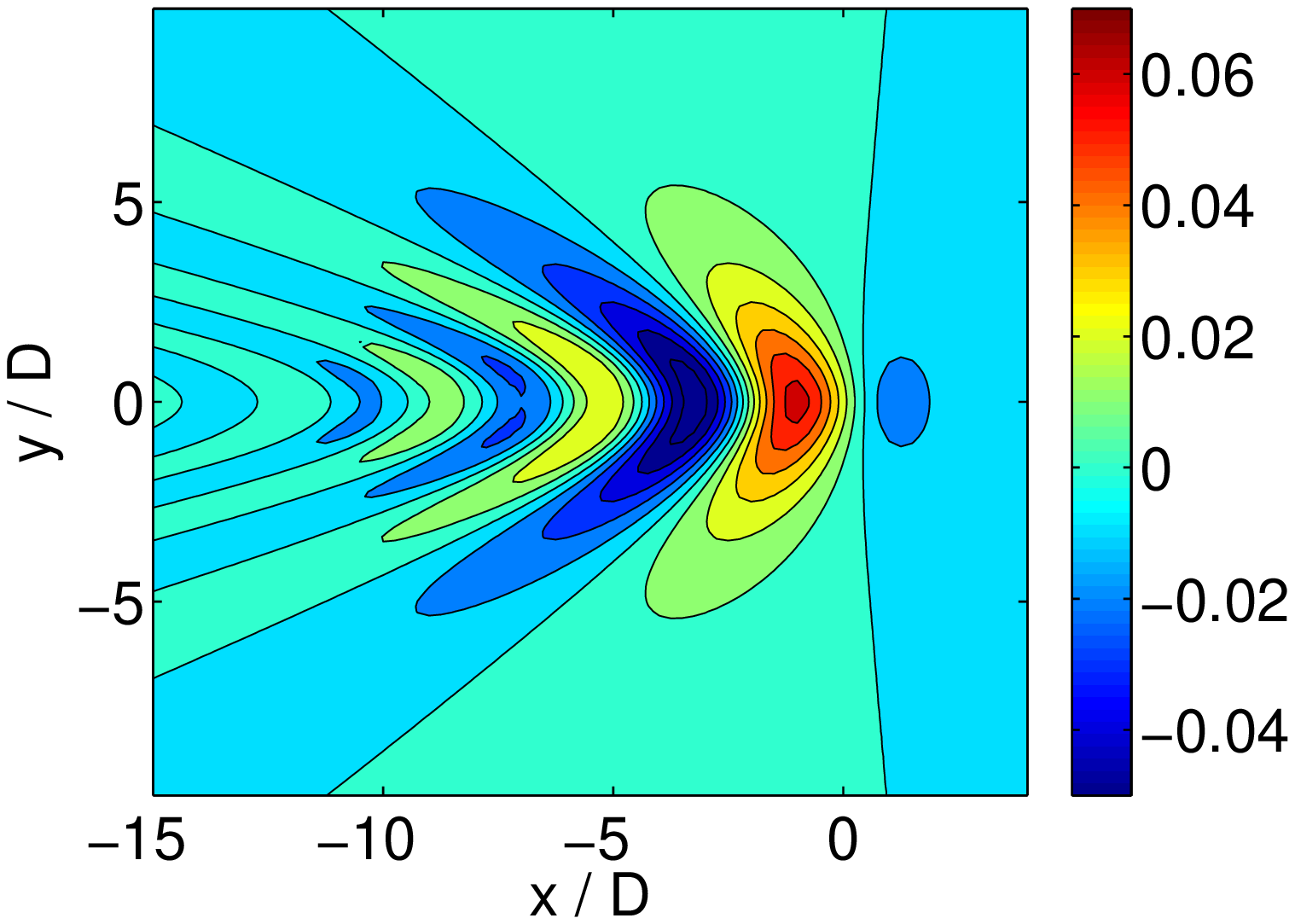}
   \end{center}
   \caption{\label{fr1} A comparison for $Fr = 1$ from the numerical simulations (upper panel)  
            and from linear theory (lower panel), on the $(x,y)$ plane located at $z = 2D$.  The plotted variable is $u/U$, 
            the $x$-component of the velocity of the wavefield.}
\end{figure}

\begin{figure}
   \begin{center}
     \includegraphics[width=\linewidth]{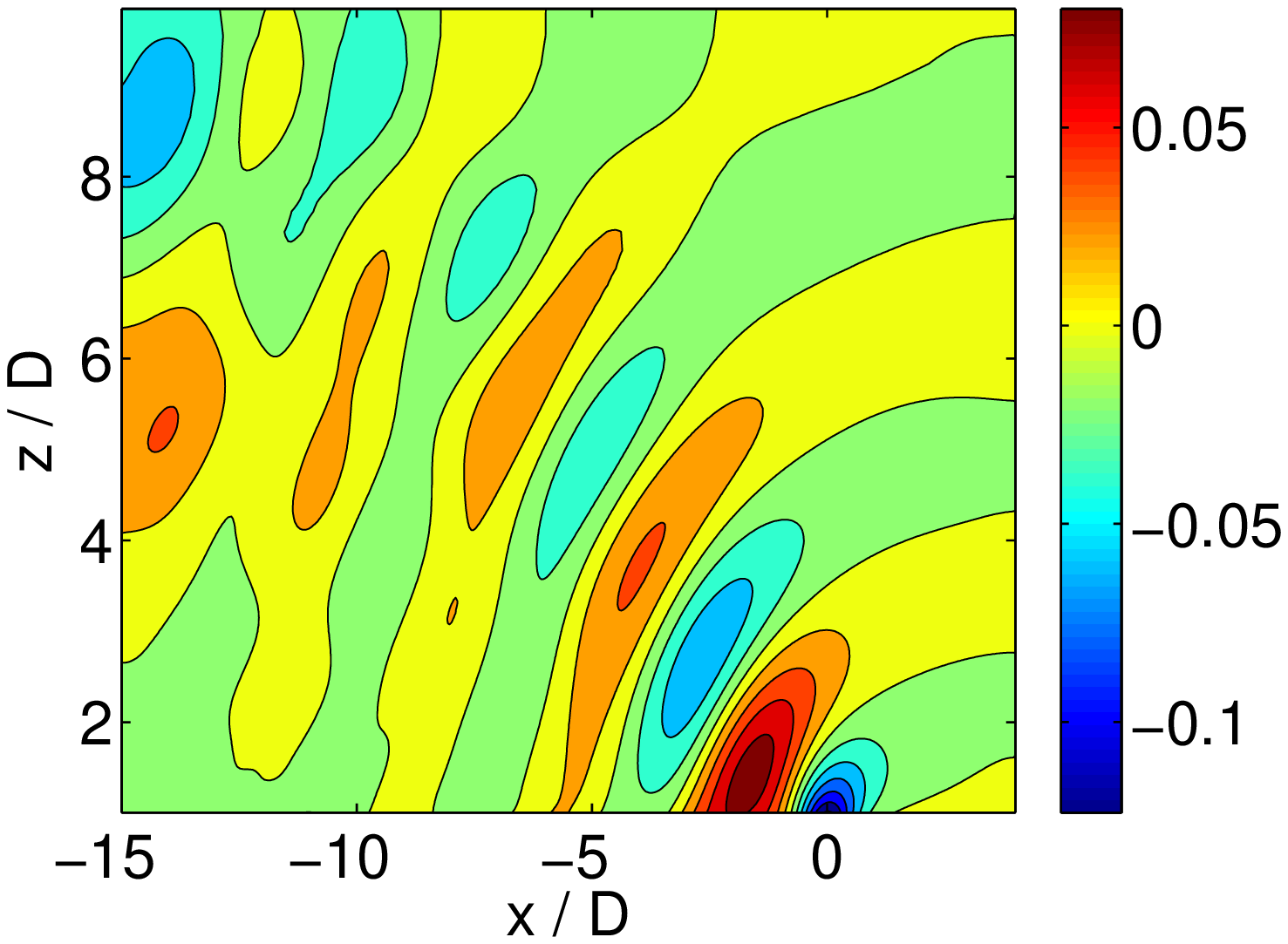}
     \includegraphics[width=\linewidth]{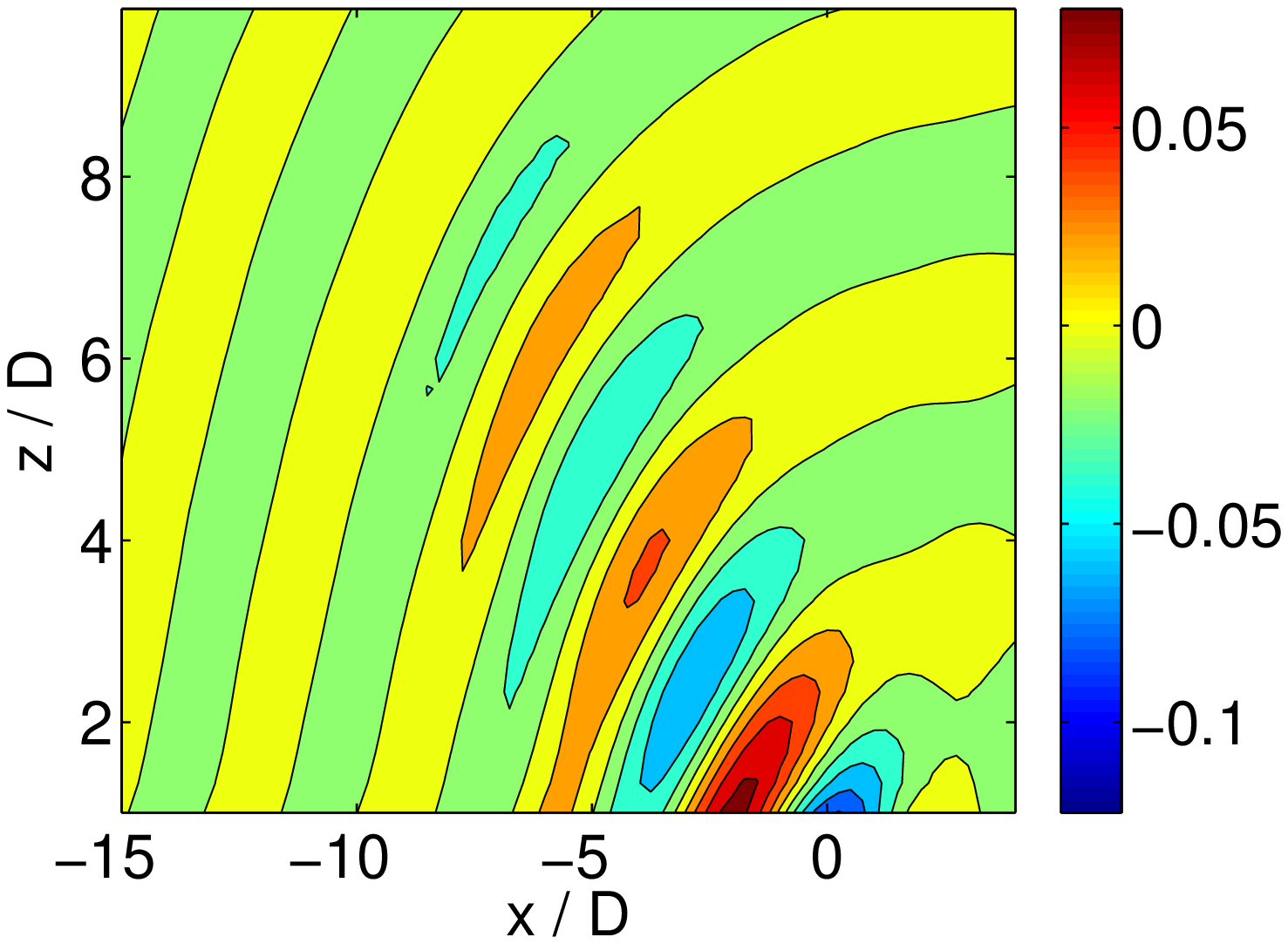}
   \end{center}
   \caption{\label{fr1_closeup} A comparison for $Fr = 1$ from the numerical simulations (upper panel)  
            and from linear theory (lower panel), for the positive$-y$ region of $(x,y)$ plane located at $z = 2D$.  The plotted variable is $u/U$, 
            the $x$-component of the velocity of the wavefield.}
\end{figure}

Figure \ref{fr1} show results for the case of $Fr = 1$. The numerical
simulation result is shown in the top panel, and the linearized theory,
with $\sigma = 0$, in the lower panel. In this case the upper and lower
boundaries of the simulations are sufficiently far away for there to be
no reflections that would be seen in the domain shown in these figures.
The two results are very close, with some small differences appearing
downstream of $x/D = -5$ near $y = 0$. At this relatively low Froude
number the body-generated waves dominate over the wake-generated waves.
It appears from these results that the proposed source representation of
the sphere produces an accurate internal wave field, even though the
source function we have used, (\ref{msource}), is most accurate for the
representation of a solid sphere only in the limit of high Froude
number. However, the discrepancies that are seen further downstream are
due to weak internal waves generated by the wake turbulence that can be
seen in figure~\ref{fr14}. This effect is better visualized in
figure~\ref{fr1_closeup}, which is a plot of $u/U$ in the vertical plane
located at $y = 0$. Only the positive $z$ half of the plane is shown. In
these plots the linear theory (lower panel) shows waves emanating only
from the sphere, where as the numerical simulations clearly show waves
emanating from the sphere and from locations downstream of the sphere.

\begin{figure}
   \begin{center}
     \includegraphics[width=0.9\linewidth]{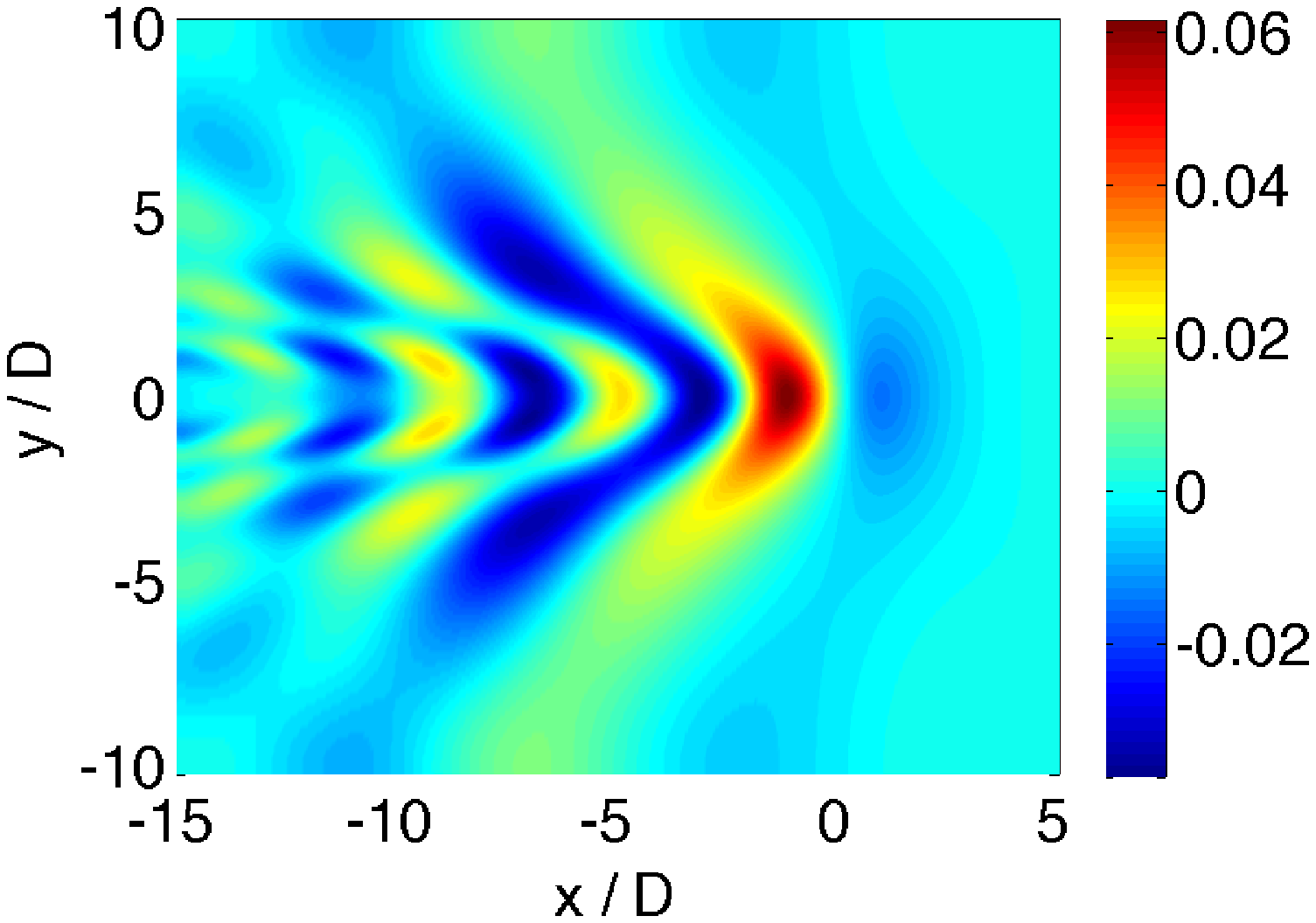}
     \includegraphics[width=0.9\linewidth]{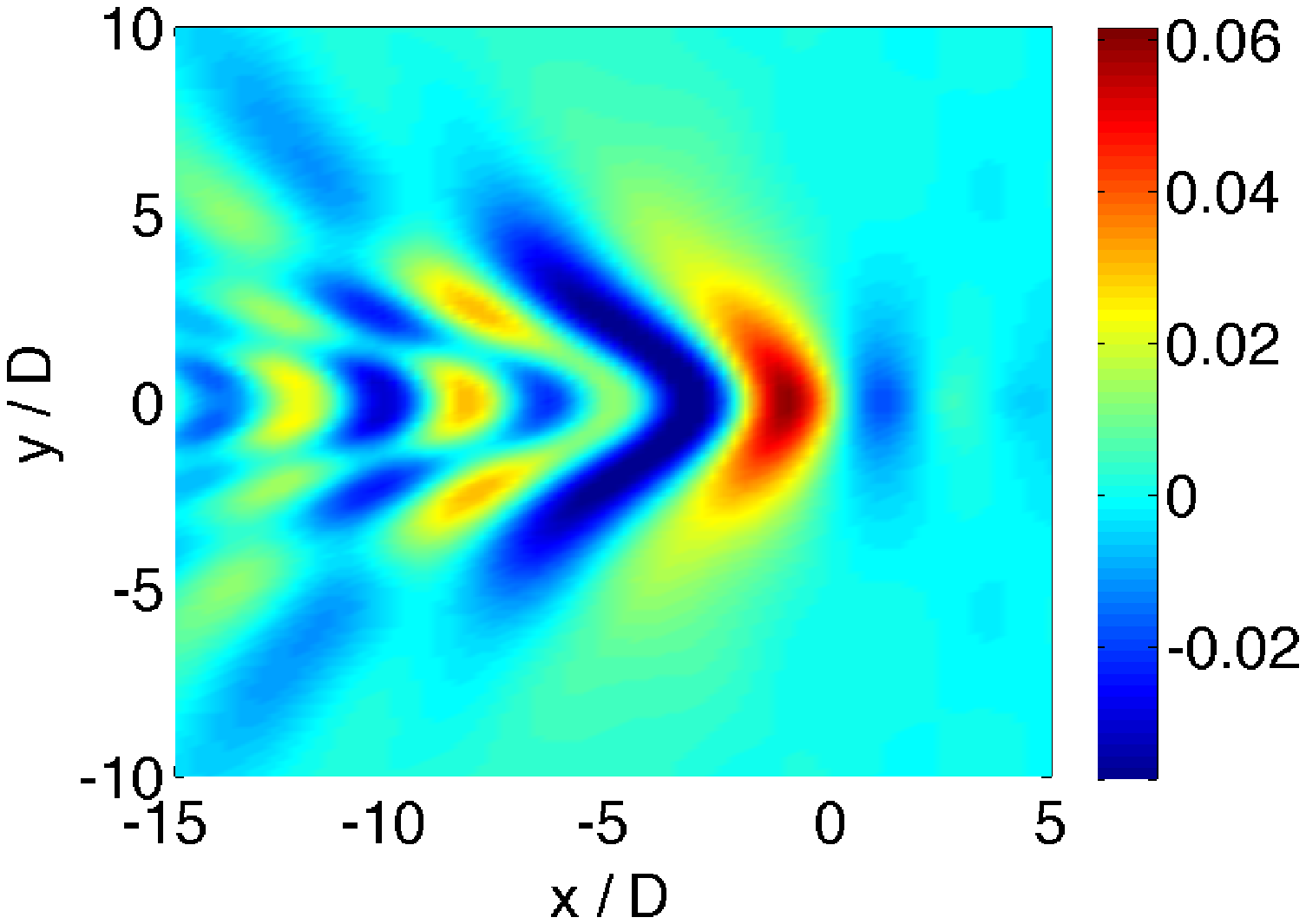}
   \end{center}
   \caption{\label{fr1_reflections} A comparison for $Fr = 1$, with upper and lower horizontal boundaries located at $z = \pm 5D$, from the numerical simulations (upper panel) and from linear theory (lower panel).  The plotted variable is $u/U$, 
            the $x$-component of the velocity of the wavefield.}
\end{figure}

Figure \ref{fr1_reflections} show results for the case of $Fr = 1$, now with the upper and lower boundaries at $z = \pm 5D$, which are close enough for the reflected waves to affect the observed wavefield. The numerical simulation result is shown in the top panel, and the linearized theory, with $\sigma = 0$, in the lower panel. Again, the two results are very close, with some small differences appearing downstream of $x/D = -5$ near $y = 0$, although the numerical simulations show some reflections from the sidewall boundaries. The linear theory was computed without sidewall boundaries. The discrepancies seen downstream and near the centerline are attributable to wake-generated waves, which are not accounted for in the linear theory.

\begin{figure}
   \begin{center}
     \includegraphics[width=\linewidth]{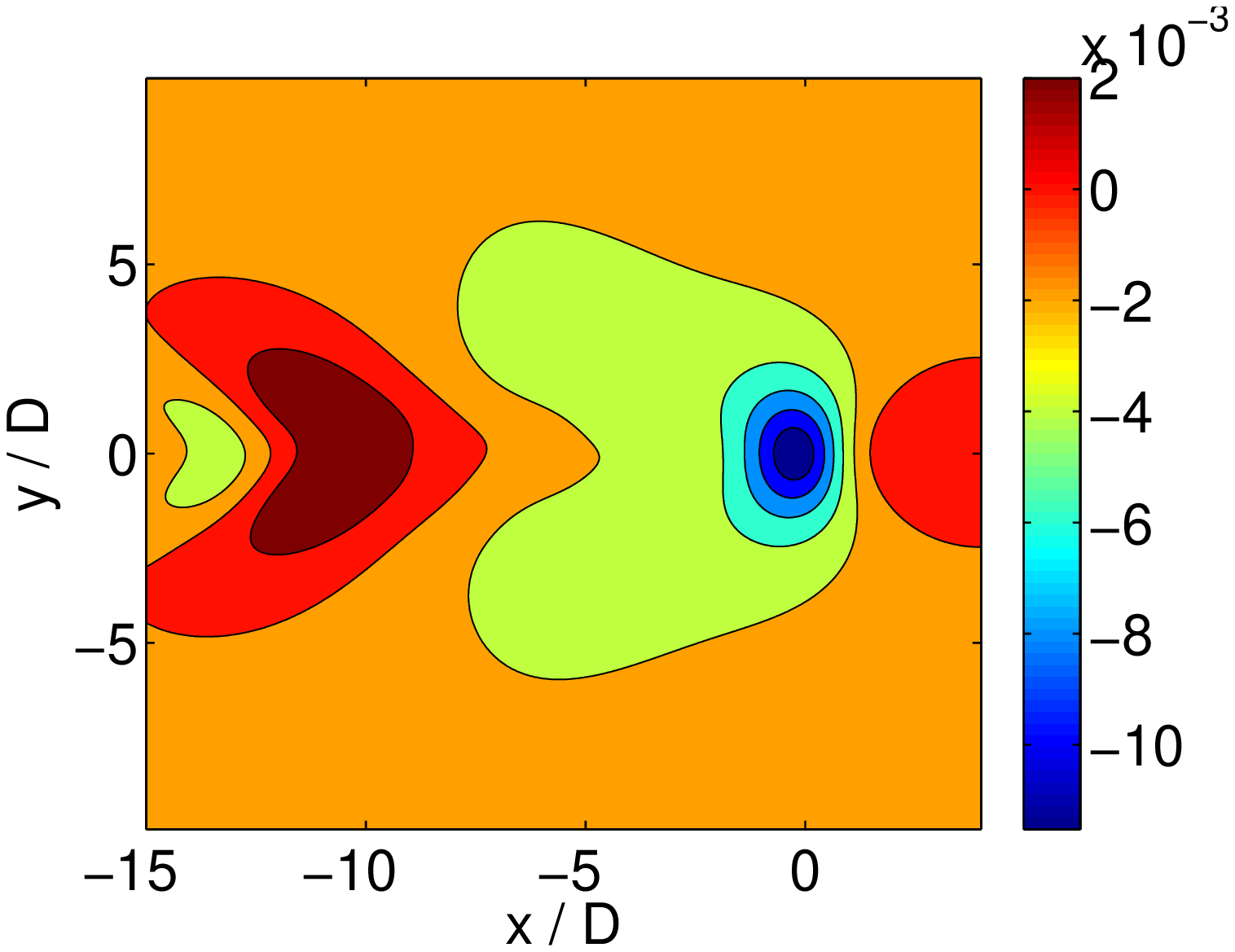}
     \includegraphics[width=\linewidth]{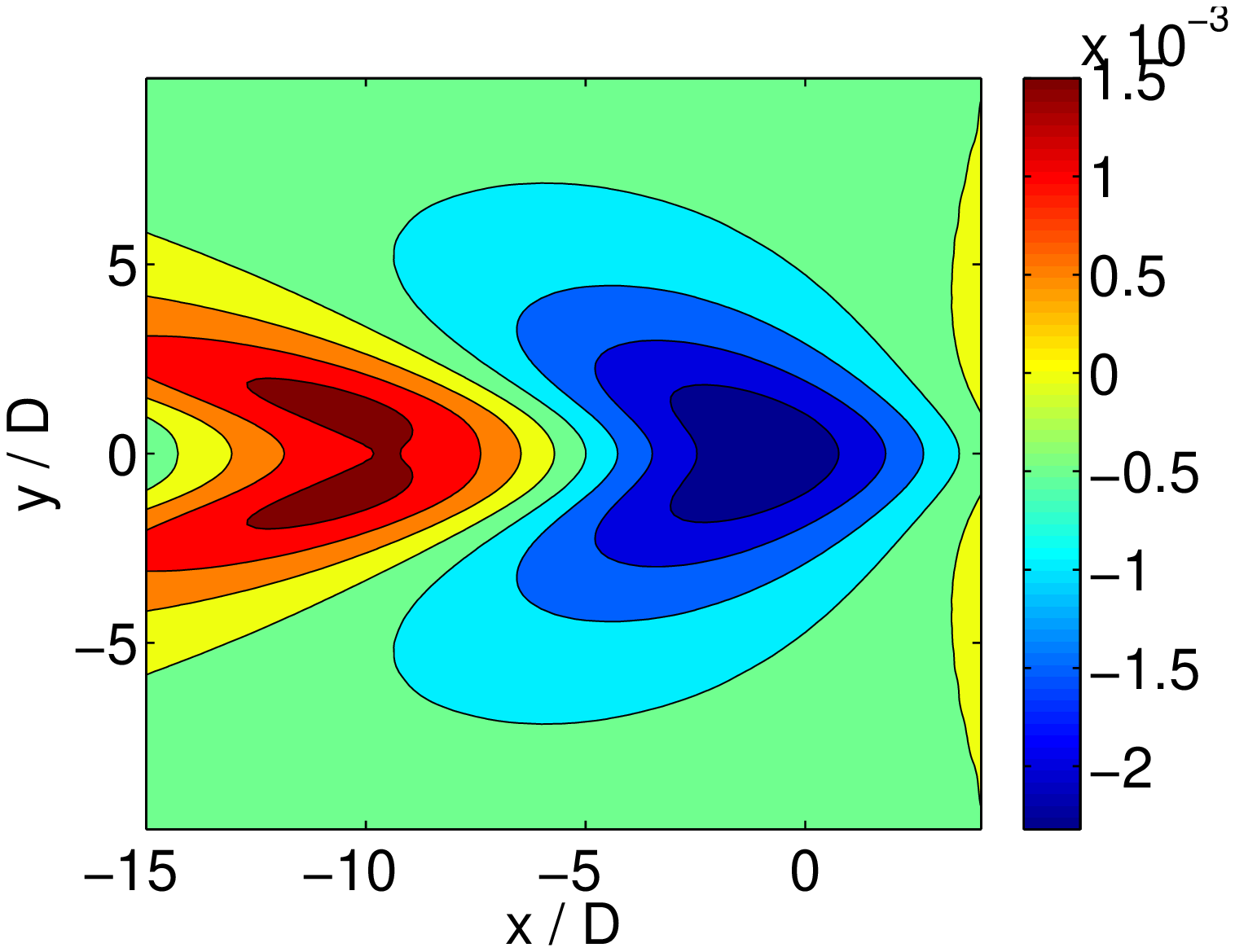}
   \end{center}
   \caption{\label{fr4} A comparison for $Fr = 4$ from the numerical simulations (upper panel) with the linear theory (lower panel),
            on the $(x,y)$ plane located at $z = 2D$. The plotted variable is $u/U$ on a horizontal plane at $z/D = 2$. 
           }
\end{figure}

\begin{figure}
   \begin{center}
     \includegraphics[width=\linewidth]{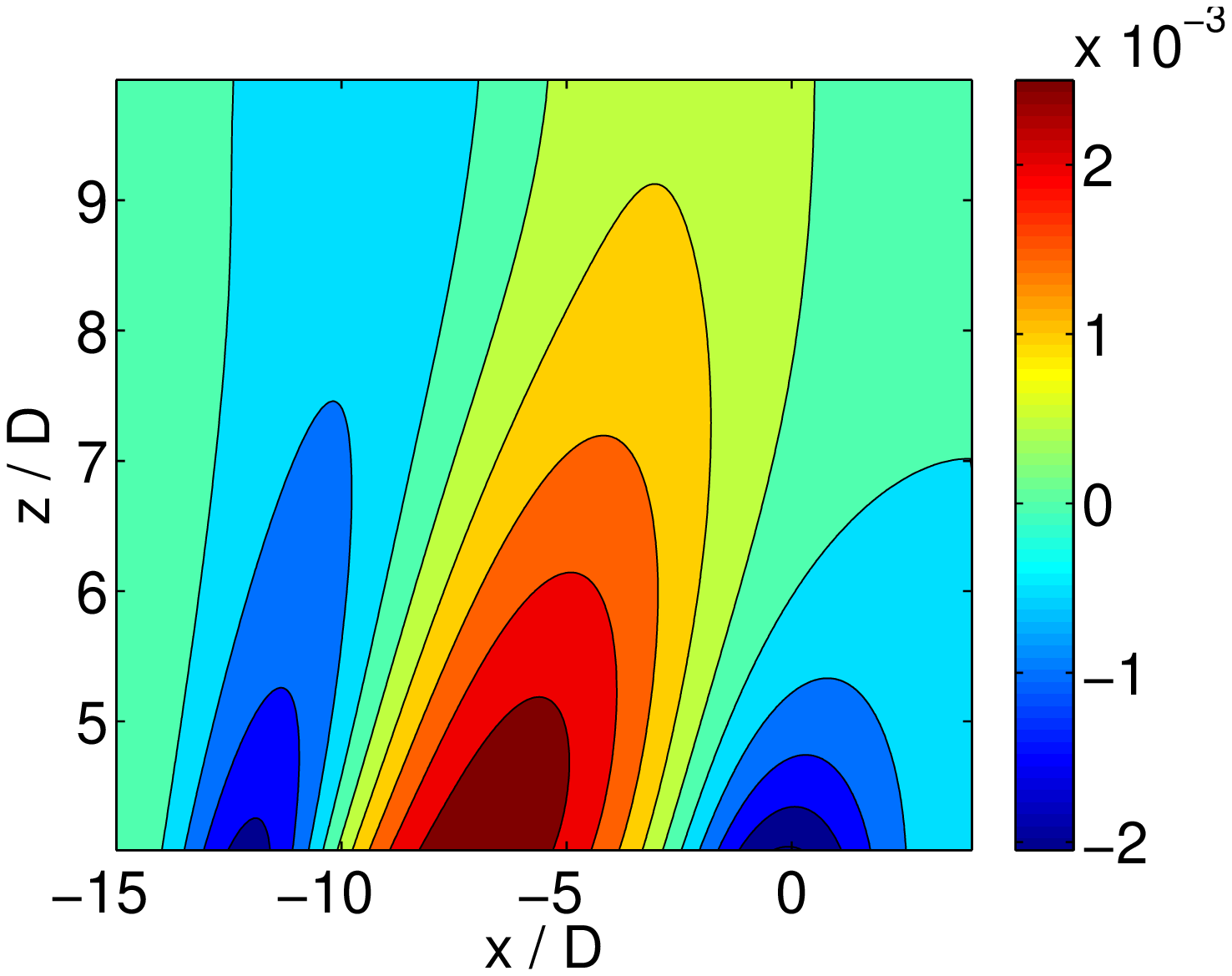}
     \includegraphics[width=\linewidth]{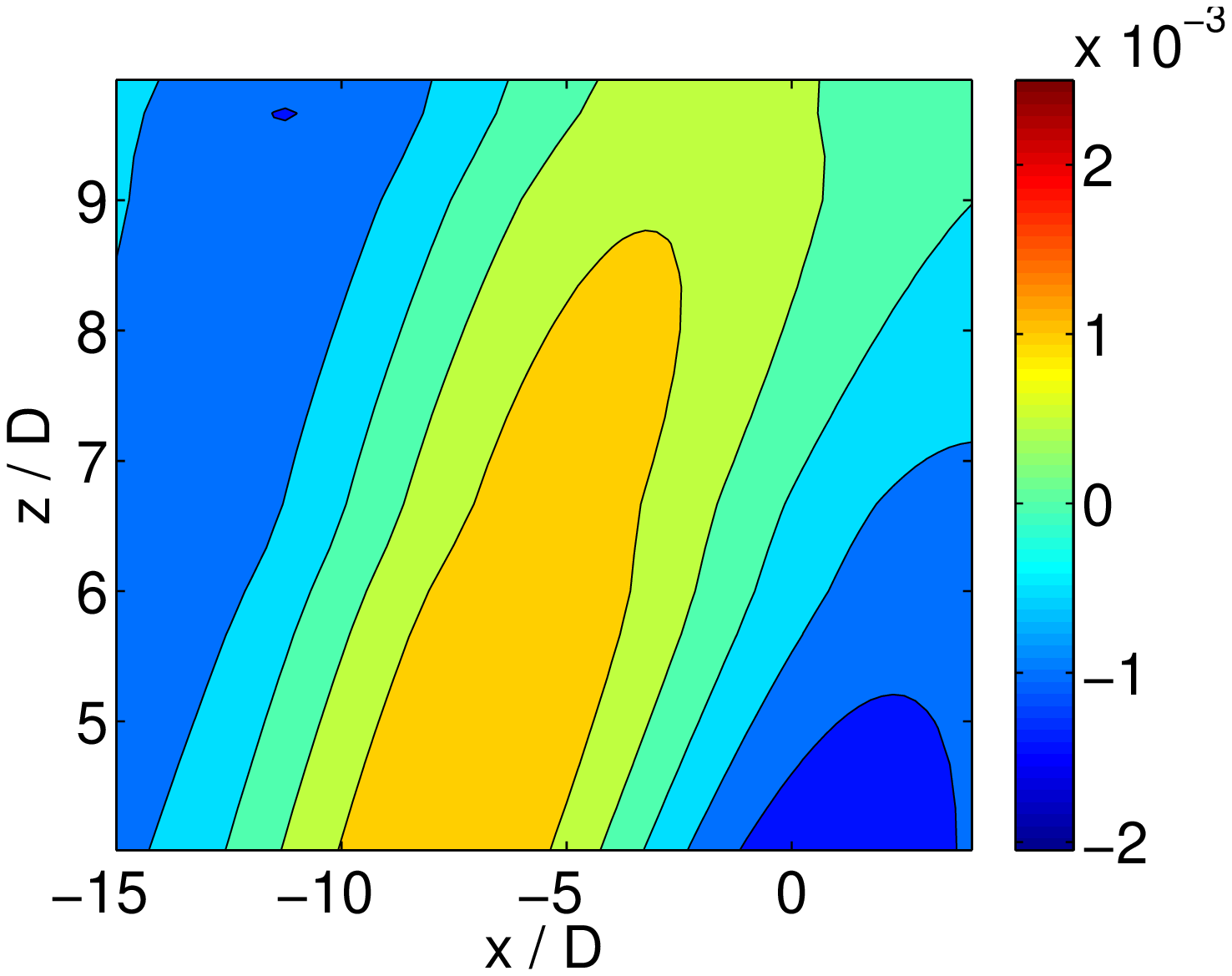}
   \end{center}
   \caption{\label{fr4_closeup} A comparison for $Fr = 4$ from the numerical simulations (upper panel) with the linear theory (lower panel), for the positive$-z$ region of the $(x,z)$ plane located at $y = 0$.
            The plotted variable is $u/U$. 
           }
\end{figure}

Figures~\ref{fr4} and \ref{fr4_closeup} show $u/U$ on a horizontal plane and a vertical plane, respectively, for $Fr = 4$.  Here the linear theory, with $\sigma = 0$, is a poor representation of the solution,  though it does give a reasonable prediction for the lateral extent of the wavefield. This is the Froude number regime in which wake-generated internal waves are expected to dominate over the body-generated waves. The wave amplitudes shown are quite small, and apparently significantly affected by the wake-generated waves. These results are very preliminary; we intend to continue the study by extending the numerical simulations farther downstream and comparing these results with the linear theory that includes combinations of nonzero source oscillation frequencies. The hope is that we will be able to determine the correct distribution of oscillation frequencies and oscillation amplitudes so that the linear theory will accurately reproduce the internal wave fields in these simulations.

\section{CONCLUSIONS} \label{sec:conclusions}

In this paper, we have extended the capabilities of the numerical model
NFA to  high Reynolds number flows around obstacles in a stratified
fluid.  We have demonstrated that NFA (1) is
capable of accurately reproducing the physics associated with highly
turbulent flows; (2) is capable of accurately reproducing highly
turbulent flows which separate due to the geometry of the body
obstructing the flow; and (3)  is capable of simulating stratified flows at low and high Froude numbers.

This work is another step in seeing how well ray theory can simulate the
internal wavefield generated by stratified flow past an obstacle.  In a
previous paper, \citet{RBSM04b} we compared ray theory with laboratory
experimental results. In this paper we have begun a comparison of the
ray simulation with the numerical model results for uniform background
with reflecting upper and lower boundaries.  The results show that the
source distribution produces a very good representation of the internal
wave field for low Froude numbers near unity where we would expect waves
generated by the body itself to dominate. At higher Froude numbers,
where the dominated wave generation is by the turbulent wake, the
comparisons are not as good. This study will be continued in the future,
now that NFA is available for these kinds of comparisons, to determine
specifically what distribution of oscillating sources best models the
wave generation by eddies in the wake.
\section{Acknowledgements}
This research was sponsored by Dr. Ron Joslin at the Office of Naval
Research (contract number N00014-08-C-0508), Dr. Tom C. Fu at the Naval
Surface Warfare Center, Carderock Division, and SAIC's Research and
Devlopment program.  The numerical simulations were supported in part by
a grant from the Department of Defense High Performance Computing
Modernization Program (\mbox{http://www.hpcmo.hpc.mil/}).  The
numerical simulations were performed on the SGI Altix ICE 8200LX at the U.S. Army
Engineering Research and Development Center.

Animated versions of the numerical simulations described here as well as
many others are available online at:
\mbox{http://www.youtube.com/waveanimations}

%
%
\bibliographystyle{28onr}
\bibliography{28onr}
\end{document}